\documentclass[10pt,journal,compsoc]{IEEEtran}
\ifCLASSOPTIONcompsoc
  \usepackage[nocompress]{cite}
\else
  \usepackage{cite}
\fi

\usepackage{times,amsmath,epsfig}
\usepackage{array,graphicx}
\usepackage{subfigure}
\usepackage{diagbox}
\usepackage{amssymb}
\usepackage[thmmarks,amsmath]{ntheorem}
\usepackage{algorithm}
\usepackage{algorithmicx,algpseudocode}
\usepackage{multirow}
\usepackage{xcolor}
\usepackage{url}
\usepackage{soul}
\usepackage{xspace}
\usepackage{longtable}
\usepackage{makecell}
\usepackage{diagbox}
\usepackage{booktabs}

\usepackage{flushend}

\usepackage[american]{babel}
\usepackage{microtype}
\usepackage{xspace}

\newcommand{\eg}{\emph{e.g.},\xspace}
\newcommand{\ie}{\emph{i.e.},\xspace}

\newcommand\figref[1]{Fig.~\ref{#1}}

\newcommand\tabref[1]{Table~\ref{#1}}

\newcommand\secref[1]{Sec.~\ref{#1}}

\newtheorem{defn}{Definition}
\newtheorem{exmp}{EXAMPLE}

\def\LL{Hourglass\xspace}
\def\MK{Piggyback\xspace}
\def\ll{HG\xspace}
\def\mk{PB\xspace}

\iftrue
\newcommand{\liliang}[1]{\large{\textbf{#1}}\normalsize}

\else
\newcommand{\liliang}[1]{#1}

\fi

\renewcommand{\baselinestretch}{0.96}
\begin{document}
\title{A Comparative Study of Consistent Snapshot Algorithms for Main-Memory Database Systems}

\author{Liang~Li, Guoren~Wang, Gang~Wu, Ye~Yuan, Lei~Chen, and Xiang~Lian, Member, IEEE
\IEEEcompsocitemizethanks{\IEEEcompsocthanksitem Liang Li, Ye Yuan are with the Department of Computer Science, Northeastern University of China, 100819.\protect\\
E-mail: \{liliang@stumail,yuanye@ise\}.neu.edu.cn
\IEEEcompsocthanksitem Guoren~Wang is with the Department of Computer Science, Beging Institute of Technology, China, CN, 100081.\protect\\
E-mail: wanggr-bit@126.com
\IEEEcompsocthanksitem Gang Wu is with the Department of Computer Science, Northeastern University of China, 100819 and with the Department of State Key Lab. for Novel Software Technology, Nanjing University, P.R. China.\protect\\
E-mail:wugang@mail.neu.edu.cn
\IEEEcompsocthanksitem Lei Chen is with the Department of Computer Science and Engineering, 
Hong Kong University of Science and Technology, Hong Kong.\protect\\
E-mail: leichen@cse.ust.hk
\IEEEcompsocthanksitem Xiang Lian is with the Department of Computer Science, Kent State University, USA.\protect\\
E-mail:xlian@kent.edu}% <-this % stops an unwanted space
%\thanks{Manuscript received XX XX, XXXX.}
}


\markboth{Journal of \LaTeX\ Class Files,~Vol.~XX, No.~X, XX~XXXX}%
{Shell \MakeLowercase{\textit{et al.}}: Bare Demo of IEEEtran.cls for Computer Society Journals}



\IEEEtitleabstractindextext{%
\begin{abstract}
In-memory databases (IMDBs) are gaining increasing popularity in big data applications, where clients commit updates intensively.
%\revise{Consistent snapshot is a key step in backup and recovery of IMDBs, thus an important factor for system performance of IMDBs.}
Specifically, it is necessary for IMDBs to have efficient snapshot performance to support certain special applications (e.g., consistent checkpoint, HTAP).
Formally, the in-memory consistent snapshot problem refers to taking an in-memory consistent time-in-point snapshot with the constraints that 1) clients can read the latest data items and 2) any data item in the snapshot should not be overwritten.
Various snapshot algorithms have been proposed in academia to trade off throughput and latency, but industrial IMDBs such as Redis adhere to the simple fork algorithm.
To understand this phenomenon, we conduct comprehensive performance evaluations on mainstream snapshot algorithms.
Surprisingly, we observe that the simple fork algorithm indeed outperforms the state-of-the-arts in update-intensive workload scenarios.
On this basis, we identify the drawbacks of existing research and propose two lightweight improvements.
Extensive evaluations on synthetic data and Redis show that our lightweight improvements yield better performance than fork, the current industrial standard, and the representative snapshot algorithms from academia.
Finally, we have opensourced the implementation of all the above snapshot algorithms so that practitioners are able to benchmark the performance of each algorithm and select proper methods for different application scenarios.
\end{abstract}
}

\maketitle

\IEEEdisplaynontitleabstractindextext

% \fi
\IEEEpeerreviewmaketitle

\IEEEraisesectionheading{\section{Introduction}\label{sec:introduction}}
\IEEEPARstart{I}{n-memory} databases (IMDBs)~\cite{zhang2015memory} have been widely adopted in various applications as the back-end servers, such as e-commerce OLTP services, massive multiple online games~\cite{Salles.12}, electronic trading systems (ETS) and so on.
For these applications, it is common to support both intensively committed updates and efficient consistent snapshot maintenance. 
Here, we use “in-memory consistent snapshot” to emphasize taking an in-memory consistent time-in-point snapshot with the constraints that 
(1) clients can read the latest data items, and
(2) any data item in the snapshot should not be overwritten.
In-memory consistent snapshot can be applied in diverse real-life applications.
Representative examples include but are not limited to the following.

\begin{itemize}
\item \textbf{Hybrid Transactional/Analytical Processing Systems (HTAP):}
Hybrid OLTP\&OLAP in-memory systems are gaining increasing popularity~\cite{OzcanTT17,stonebraker2007end,kemper2011hyper,plattner2009common,lang2016data,funke2011benchmarking,Meng2017,farber2012sap,sikka2013sap}.
In traditional disk-resident database systems, the OLTP system needs to extract and transform data to the OLAP system. That is, OLTP and OLAP are usually separated in two systems.
Due to the high performance of in-memory database systems, it becomes viable to exploit OLTP snapshot data as an OLAP task and build a hybrid system.
In fact, database vendors including Hyper~\cite{kemper2011hyper}, SAP HANA~\cite{farber2012sap,sikka2013sap} and SwingDB~\cite{Meng2017} have already applied in-memory snapshot algorithms in Hybrid Transactional/Analytical Processing Systems.
\item \textbf{Consistent Checkpoint:}
System failures are intolerable in many business systems.
For instance, Facebook was out of service for approximately 2.5 hours in 2010.
There was a worldwide outage, and 2.8TB memory data were cleared~\cite{facebook}.
Consistent checkpoints are important to avoid long-time system failures and support rapid recovery; in-memory systems such as Hekaton\cite{diaconu2013hekaton} and Hyper~\cite{muhe2011efficiently} typically perform \textit{consistent checkpoint} frequently.
Checkpoint works by taking a ``consistent memory snapshot" of the runtime system and dumping the snapshot asynchronously.
The key step is to take a consistent snapshot efficiently.
Inefficient snapshot algorithms may accumulatively lead to system performance degradation and thus unacceptable user experience in update-intensive applications.
\end{itemize}

However, the unavoidable fact is that the accumulated latency brought by the
snapshot maintenance may have significant impacts on system throughput and response time.
Improper handling of snapshot may result in latency spikes and even system stalls. Thus, pursuing a fast snapshot with low and uniform
overhead, or one that is \textit{lightweight}, is the focus of in-memory
snapshot algorithms.

The wide applications of in-memory consistent snapshot have attracted the interest of academia.
Some representative snapshot algorithms are Naive Snapshot (NS)~\cite{bronevetsky2006recent,schroeder2007understanding}, Copy-on-Update (COU)~\cite{Salles.12,liedes2006siren,Cao2013Fault}, Zigzag (ZZ)~\cite{Cao.13} and PingPong (PP)~\cite{Cao.13}.
In addition, the simple fork~\cite{fork-wiki} function is used as a common snapshot algorithm in industrial systems.
However, it is often difficult for practitioners to select the appropriate in-memory snapshot algorithm due to the lack of a unified, systematic evaluation on existing snapshot algorithms.
This work is primarily motivated by this absence of performance evaluation, which is described in more detail as follows.

\subsection{Motivation}
%最后一句：流行的产业界IMDBs采用简单的fork算法是出于工程上简单化的原则？还是出于系统性能的考虑？

\textbf{1. Why do popular industrial IMDBs, \eg Redis/Hyper, utilize the simple fork() function instead of state-of-the-art snapshot algorithms?}
As mentioned above, various in-memory consistent snapshot algorithms have been proposed in academia to trade off between latency and throughput.
However, it is interesting that popular industrial IMDBs such as Redis/Hyper still apply the simple fork() function as the built-in algorithm for consistent snapshot.
It is worth investigating whether this is due to the simplicity of fork()'s engineering implementation or its good system performance (\eg high throughput and low latency).

%\textbf{2. How to take snapshot for concurrent situation? }
%以上的算法，都是基于一个假设：数据在执行快照的时刻就是一致的，这意味着我们不需要额外的去维护快照的一致性了。
%比如说，对于串行事务系统中，每个事务提交的时刻，数据本身就是一致的。
%然而，这些算法都没有去考虑并行下，维护一致性的问题。
%据我们所知，传统维护一致性快照的方式是：等待还未提交的活动事务提交并且阻塞新事务的启动，从而等待所有事务提交结束，达到一致性。
%这样做的问题在于会导致一段额外的阻塞。
\textbf{2. Are state-of-the-art snapshot algorithms inapplicable to update-intensive workload scenarios?}
Many modern in-memory applications are highly interactive and involve intensive updates.
The performance of the state-of-the-arts from academia and industry in large-scale update-intensive workload scenarios is not known.
If no existing algorithms fit, can we modify and improve the state-of-the-arts for this scenario?

\textbf{3. Can we provide unified implementation and benchmark studies for future studies?}
A frustrating aspect of snapshot algorithm research is the lack of a unified implementation for fair and reproducible performance comparisons.
Since new application scenarios are continually emerging, researchers would benefit by making unified implementation and evaluation of existing snapshot algorithms accessible to all.

\subsection{Contributions}

\begin{figure}[htb]
	\centering
	\includegraphics[width=0.40\textwidth]{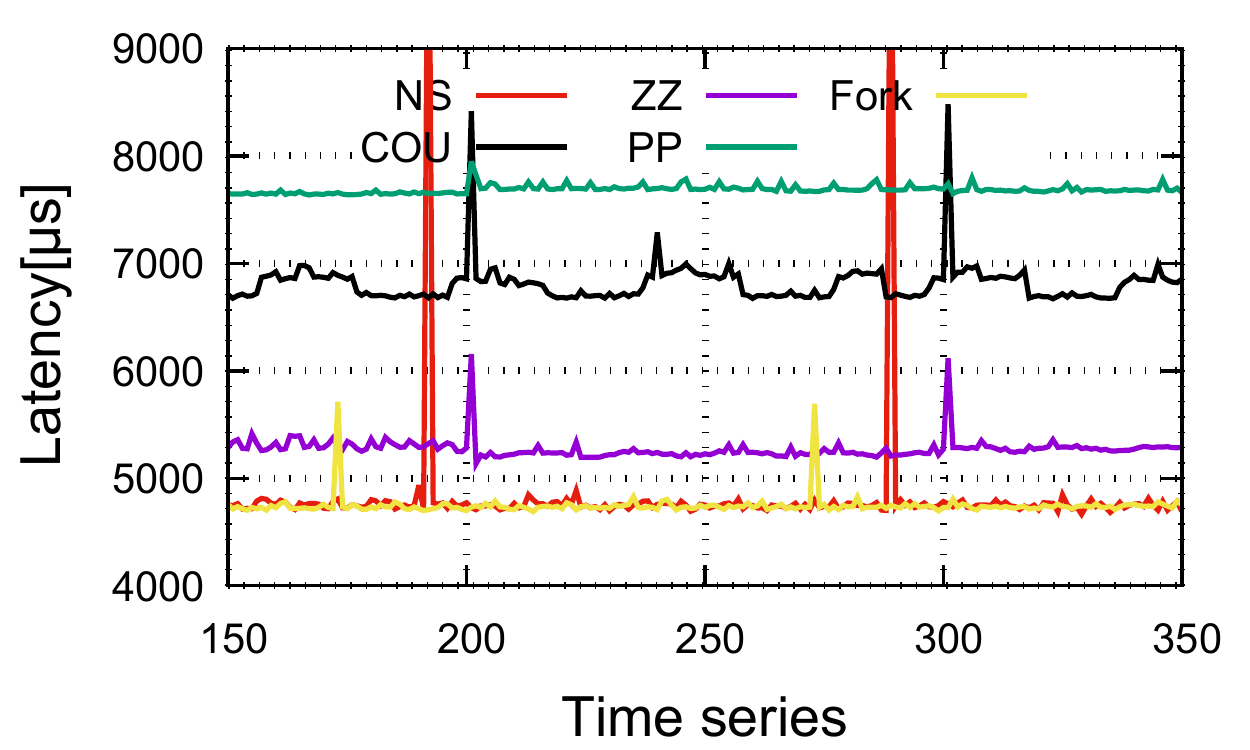}
	\caption{Comparison of State-of-art algorithms}
	\label{fig:introlatency}
\end{figure}

\textbf{1. We find that the simple fork() function indeed outperforms the state-of-the-arts in update-intensive workload scenarios.}
Snapshot algorithms for update-intensive workloads should have consistently low latency.
This requirement can be assessed by average latency and latency spikes.
We conduct large-scale experiments on five mainstream snapshot algorithms (NS, COU, ZZ, PP, Fork).
\figref{fig:introlatency} shows illustrative latency traces of five mainstream snapshot algorithms.
NS has low average latency but also high-latency spikes, meaning high latency when taking snapshots.
In contrast, PP has no latency spikes but incurs higher average latency.
Surprisingly, we observe that the simple fork algorithm indeed outperforms the remaining algorithms.
That is, fork() has low average latency and almost no high latency spikes.
These experimental results can explain why popular industrial IMDBs prefer the simple fork algorithm rather than state-of-the-art algorithms from academia. This is more programming friendly.

\textbf{2. We propose two simple yet effective modifications of the state-of-the-arts that exhibit better tradeoff among latency, throughput, complexity and scalability.}
Based on the aforementioned experiments with mainstream snapshot algorithms, we identify the drawbacks of the existing research and propose two lightweight improvements based on state-of-the-art snapshot algorithms.
In particular, extensive evaluations on synthetic data and Redis, the popular industrial IMDB, show that our lightweight improvements yield better performance than fork, the current industrial standard, and the representative snapshot algorithms of academia.
%\revise{What's more, the algorithms can be easily adapted to widely use cases (a.k.a, concurrent transaction execution scenarios) by leverage virtual snapshot techniques, with the same time keep a good performance.}
In addition, the algorithms can not only easily adapt to widely used cases but also maintain good performance with the snapshot technique.

\textbf{3. We opensource our implementations, algorithmic improvements, and benchmark studies as guidance for future researchers.}
We implement five mainstream snapshot algorithms and two improved algorithms and conduct comprehensive evaluations on synthetic datasets.
The implementations and evaluations have been released on GitHub\footnote{\url{https://github.com/bombehub/FrequentSnapshot}}.
We further integrate the two improved algorithms into Redis and investigate the scalability with the Yahoo! Cloud Serving Benchmark (YCSB)~\cite{Cooper.14}.
The implementations and evaluations are also publicly accessible\footnote{\url{https://github.com/bombehub/RedisPersistent}}.
We envision our experiences as providing valuable guidance for future snapshot algorithm design, implementation, and evaluation.

This paper is a complete description of a previous brief
version of this work~\cite{myposter}. The main additions include a number
of examples in the background and motivation, the theoretical
foundation and implementation of our algorithms, and presentation and analysis of extensive experimental results.
Furthermore, we adapt the proposed algorithms to the more general
concurrent transaction-execution case for comparison with the
CALC algorithm~\cite{ren2016low-overhead}.

%\revise{This paper is a complete description of our previous brief version of this work \cite{myposter}. The main additions include a number of examples in the background and motivation, the theoretical foundation and implementation of the proposed algorithms, and presentation and analysis of extensive experimental results. Furthermore, we adapt the proposed algorithms to the more general concurrent transaction-execution case in order to compare with the CALC algorithm \cite{ren2016low-overhead}.}

The rest of the paper is organized as follows.
In Section \ref{Sec:Preliminaries}, we define and model the problem of consistent snapshot.
Existing algorithms and two proposed algorithms are detailed in Section \ref{Sec:Algorithms}.
We discuss a more general case in Section \ref{sec:discuss}.
To show the feasibility of the algorithms, we first evaluate them with a synthetic dataset in Section \ref{sec:comparison} and then integrate them into Redis and benchmark them with YCSB in Section \ref{Sec:SystemValidations}.
We conclude in Section \ref{Sec:Conclusions}.

%\IEEEpubidadjcol

\section{Preliminaries}\label{Sec:Preliminaries}
%\subsection{Background}
%talk about HTAP and Checkpoint work.

\subsection{Problem Statement}\label{Subsec:DefinitionModel}
In this work, we compare, analyze and improve snapshot algorithms designed for in-memory databases, particularly in update-intensive scenarios.
First, we formally define the in-memory consistent snapshot problem as follows.
\begin{defn}[In-Memory Consistent Snapshot]
	Let $D$ be an update intensive in-memory database.
	A consistent snapshot is a consistent state of D at a particular time-in-point,
	which should satisfy the following two constraints:
	\begin{itemize}
		\item \textbf{Read constraint:}
		Clients should be able to read the latest data items.
		\item \textbf{Update constraint:}
		Any data item in the snapshot should not be overwritten. In other words, the snapshot must be read-only.
	\end{itemize}
\end{defn}

An in-memory consistent snapshot algorithm for update-intensive applications must fulfill the following requirements.
\begin{itemize}
	\item \textbf{Consistent and Full Snapshots.}
	``Dirty'' (\ie inconsistent) snapshots are intolerable.
	Furthermore, since we do not consider applications such as incremental backups, full snapshots that materialize all the application data states are indispensable.
	\item \textbf{Lock-free and Copy-Optimized.}
	Locking and synchronous copy operations are the main causes of snapshot overhead~\cite{Salles.12}.
	Therefore, lock-free and copy-optimized snapshot algorithms are more desirable.
	% 为了维护快照，系统会需要阻塞一定的时间。
	\item \textbf{Low Latency and No Latency Spikes.}
	Latency spikes (\ie periodic sharp surges in latency) lead to system quiescing, which degrades user experience.
	\item \textbf{Small Memory Footprint.}
	The snapshot algorithms should incur low overhead and memory to support large-scale update-intensive applications.
\end{itemize}

%\begin{itemize}
%  \item \textbf{Serial Transaction Management.}
%  In this situation, taking snapshots are relatively easy because the physical data consistency is naturally guaranteed at the end of each transaction.
%  Snapshots can be taken at the physical consistent time-points.
%  \item \textbf{Concurrent Transaction Management.}
%  In this situation, it is more complicated to take snapshots.
%  To ensure a consistent physical data state, all new coming transactions should be blocked until all the active transactions have been committed.
%\end{itemize}

\subsection{Model and Framework}
We model the in-memory dataset $D$ as a page array.
Each page contains multiple data items, and the size of each page is 4 KB as in typical operating systems.
To simplify illustration, we assume only one item per page in the running examples throughout this paper.

Interface \ref{alg:framework} shows the snapshot algorithm framework.
We assume two kinds of threads: the \textit{client} thread and the \textit{snapshotter} thread.
The client thread continuously performs large amounts of \textbf{Read()} and \textbf{Write()} function requests, as in update-intensive applications.
The snapshotter thread is responsible for taking snapshots periodically.
\textbf{Trigger()} is periodically called to check if the previous snapshot process has completed.
If yes, it invokes \textbf{TakeSnapshot()}.
Then, \textbf{TraverseSnapshot()} is invoked to traverse the generated snapshot.
Each interval (a.k.a, \textit{period}) consists of two phases, \ie the \textit{taken phase} and the \textit{access phase}.
It should claim that \textbf{Trigger()} is always invoked at a physical consistent time point when no transactions are uncommitted.

\begin{algorithm}[!htb]
	\floatname{algorithm}{Interface}
	\caption{ Snapshot Algorithm Framework.}
	\label{alg:framework}
	\begin{algorithmic}[1]		
		\State Client::Read(index);
		\State Client::Write(index,newValue);		
		\State Snapshotter::Trigger();
		\State Snapshotter::TakeSnapshot();
		\State Snapshotter::TraverseSnapshot();
	\end{algorithmic}
\end{algorithm}

%As next, we illustrate the process of snapshot taking in the serial situation by an example.
%We will discuss the more complex concurrent situation examples in \secref{sec:cc}.

In the rest of this paper, we illustrate the process of representative snapshot algorithms via examples, and we start with an example of Naive Snapshot.

%We choose checkpoint as the application to illustrate the problem.
%Now we can model the problem of consistent checkpointing as follows.
%\textbf{模型是什么样的？以及NS的例子}
%串并行对快照执行有一定影响，在这里我们以串行模型展开讨论，后面的章节会对并行的情况进行讨论。

\begin{figure}[htb]
	\centering
	\includegraphics[width=0.40\textwidth]{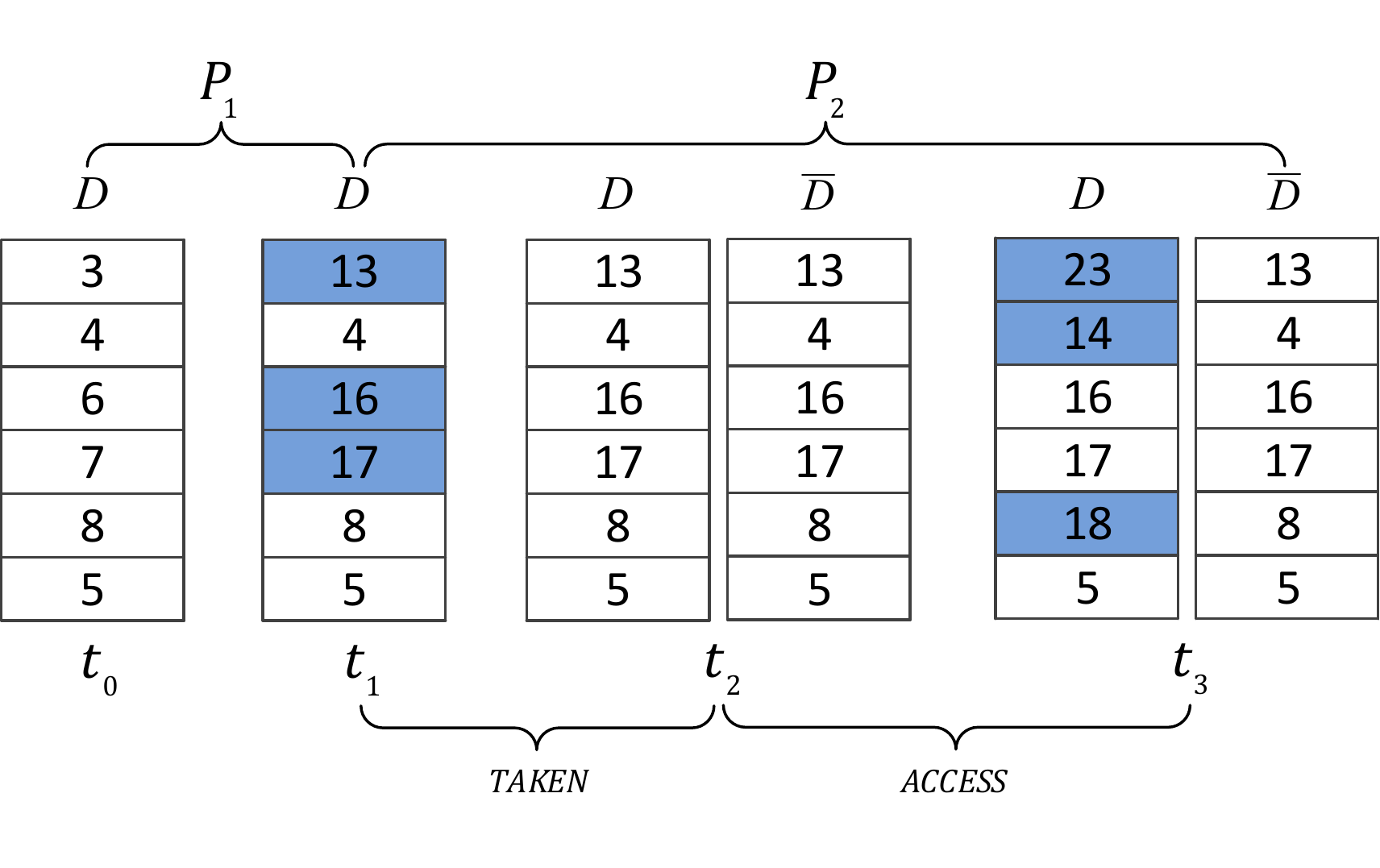}\\
	\caption{Running example for Naive Snapshot (NS)}
	\label{fig:ns}
\end{figure}

\begin{table}[htbp]
	\centering
	\caption{Transaction Data}
	\begin{small}
		\begin{tabular}{|c|c|l|}
			\hline
			{\bfseries Period} &   {\bfseries Transactions}  & {\bfseries Data to be Updated} \\
			\hline
			$ P_1 $ &   $ T_1 $    & $ <0, 13> $ \\
			\cline{2-3}
			&   $ T_2 $    & $ <2, 16> $, $ <3, 17> $ \\
			\hline
			$ P_2 $ &   $ T_3 $    & $ <0, 23> $  \\
			\cline{2-3}
			&   $ T_4 $    & $ <1, 14> $, $ <4, 18> $ \\
			\hline
		\end{tabular}
	\end{small}
	\label{tab:dataexmp}%
\end{table}%

\begin{exmp}[Naive Snapshot]\label{eg:ns}
	Assume an initial dataset $D$ = $\{3, 4, 6, 7, 8, 5\}$ at time $t_0$.
	\tabref{tab:dataexmp} shows the client data streams to be updated, and \figref{fig:ns} shows the data state.
	We further assume periodic snapshot taking.
	In the first period $P_1$ $(t_0 \rightarrow t_1)$, there are two transactions $T_1$ and $T_2$, where each update is represented by an $<index, value>$ pair.
	At the end of $P_1$ (\ie at time $t_1$), the updated data state $D$ = $\{13, 4, 16, 17, 8, 5\}$.
	
	We need to take a snapshot of $D$ at time $t_1$.
	First, the client is blocked during the snapshot taken phase ($t_1 \rightarrow t_2$),  and the snapshotter thread duplicates and bulk copies all the data $D$ to snapshot $\overline{D}$.
	Next, in the access phase ($t_2 \rightarrow t_3$), the client thread writes $T_3$ and $T_4$ to $D$, and the snapshotter thread can access the snapshot from $\overline D$.
	Note that the client can read the latest data from $D$ during the entire period.	
\end{exmp}

\section{In-Memory Consistent Snapshot Algorithms}\label{Sec:Algorithms}
In this section, we review the mainstream snapshot algorithms for in-memory database systems.
Based on in-depth analysis on the drawbacks of existing algorithms, we also propose modifications and improvements of existing snapshot algorithms.

\subsection{Representative Snapshot Algorithms}\label{SubSec:ExistingAlgorithms}
This subsection describes four mainstream snapshot algorithms (NS, COU, ZZ, PP) proposed by academia.
%Note that the fork function adopted in industrial IMDBs is a variant of Copy-on-Update (COU), which will be explained later.
\subsubsection{Naive Snapshot}
Naive snapshot (NS)~\cite{bronevetsky2006recent}\cite{schroeder2007understanding} takes a snapshot of data state $D$ during the taken phase when the client thread is blocked.
Once the snapshot $\overline D$ is taken in memory, the client thread is then resumed.
Meanwhile, the snapshotter thread can access or traverse the snapshot data $\overline D$ asynchronously.
Clients can read the latest data from $D$ during the entire process.
EXAMPLE~\ref{eg:ns} shows an example.

\subsubsection{Copy-on-Update and Fork}
Copy on Update (COU)~\cite{Cao2013Fault} utilizes an auxiliary data structure $\overline D$ to shadow copy $D$ and a bit array $\overline D_b$ for recording the page update states of $D$.
Any client write on a page of $D$ for the first time leads to a shadow page copy to the corresponding page of $\overline D$ and a setting to the corresponding bit of $\overline D_b$ to indicate the state before the page update.
In COU, the snapshotter thread can utilize the $\overline D_b$ to access the snapshot.
We refer readers to~\cite{Cao2013Fault} for more details.

Note that COU has many variants~\cite{Salles.12}\cite{liedes2006siren}, and here, we refer to the latency-spike-free implementation in~\cite{Cao2013Fault}.
The fork function~\cite{fork-wiki} is also a system-level COU variant.
Many popular industrial systems such as Redis~\cite{redis-web} and Hyper~\cite{kemper2011hyper} exploit fork to take snapshots.

%In engineering, there exist more COU variants.
%Liedes et al.\ proposed a tuple shadowing copy algorithm called SIREN \cite{liedes2006siren} which is an end-to-end COU variant.
%Though it is memory compact, more overheads are required on fine-grained locks and logical page maintains.
%In addition, if CALC runs in the serial mode, it can also be seen as a COU.

\begin{figure}[htb]
	\centering
	\includegraphics[width=0.48\textwidth]{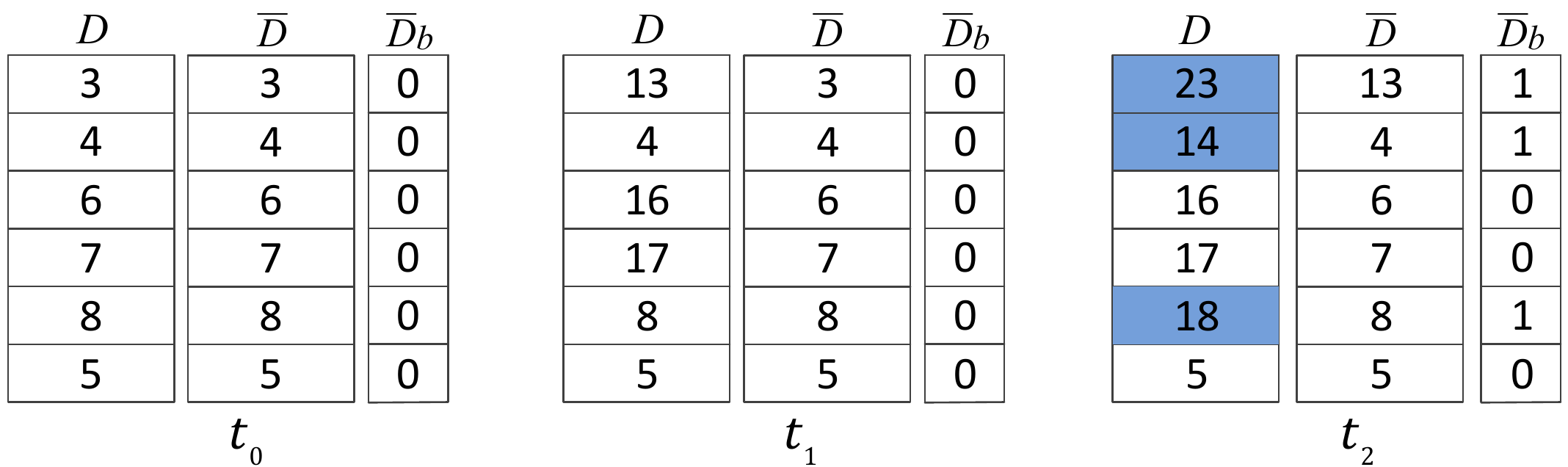}\\
	\caption{Running example for Copy on Update (COU)}
	\label{fig:cou}
\end{figure}

\begin{exmp}[Copy-on-Update]
As with EXAMPLE~\ref{eg:ns}, dataset $D$ update from $\{3,4,6,7,8,5\}$ to $\{13,4,16,17,8,5\}$.
To take a snapshot at time $t_1$, the incoming transaction $T_3$ and $T_4$ should not overwrite the snapshot data.
So, COU copies the snapshot data to the shadow data $\overline D$ and sets the bit flag $\overline D_b$ to keep track of the ``dirty'' data.
The snapshotter thread is able to access the snapshot data using the bit flag, as shown in \figref{fig:cou}.
However, there must be exclusive locks between the client thread and the snapshotter thread, which leads to performance loss.
\end{exmp}

\subsubsection{Zigzag}

Zigzag (ZZ)~\cite{Cao.13} employs one shadow copy $\overline D$ (of the same size as $D$) and two auxiliary bit arrays $\overline D_{br}$ and $\overline D_{bw}$.
For a page $i$, $\overline D_{br}[i]$ and $\overline D_{bw}[i]$ are responsible for indicating which copy the client should read from or write to, respectively.
Hence, $\neg \overline D_{bw}[i]$ indicates the copy the snapshotter thread should access since this copy cannot be written by the client.
ZZ is able to avoid being overwritten and retain untouched snapshot data with the help of $\overline D_{bw}[i]$.

\begin{figure}[htb]
	\centering
	\includegraphics[width=0.48\textwidth]{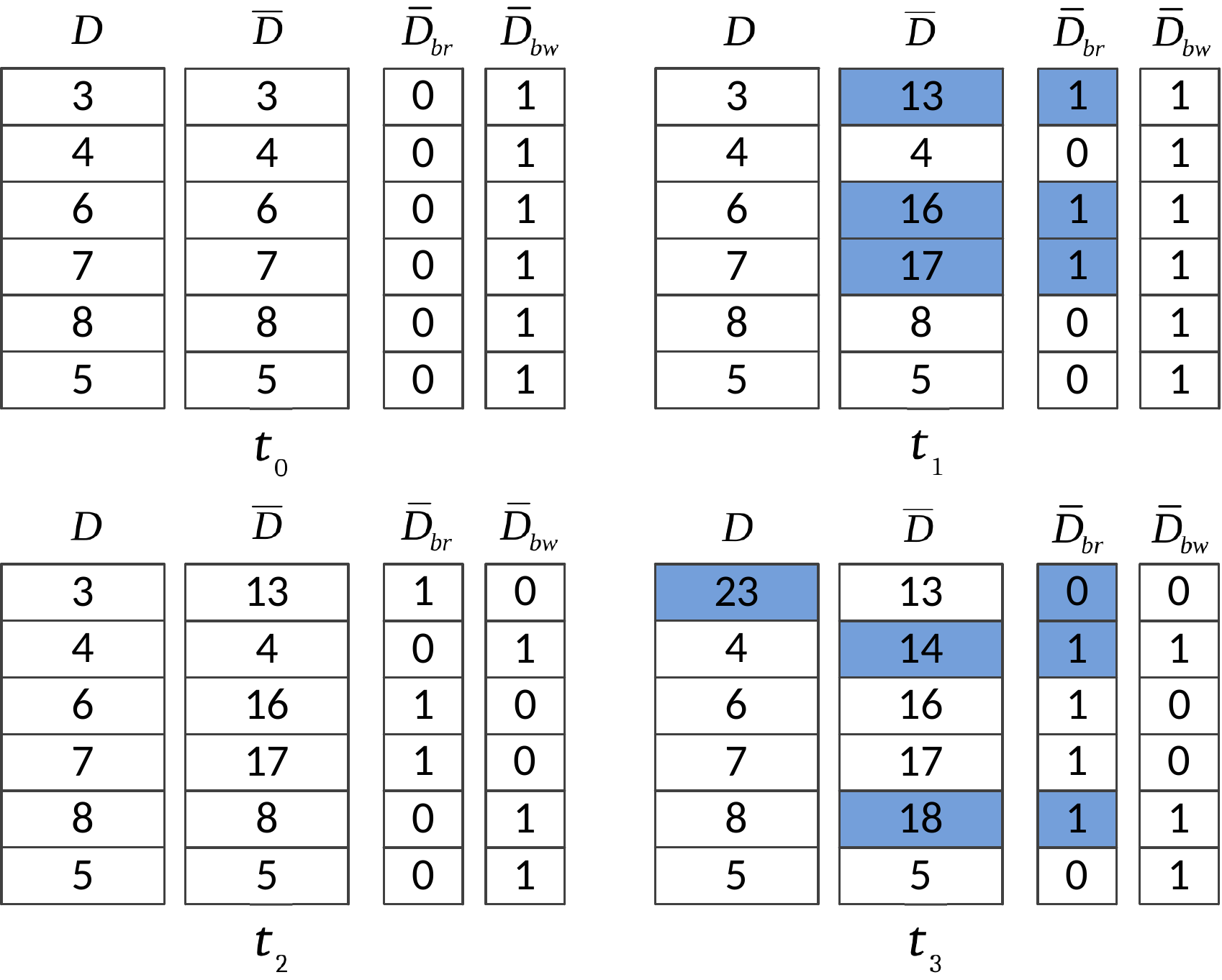}\\
	\caption{Running example for Zigzag (ZZ)}
	\label{fig:zz}
\end{figure}

\begin{exmp}[Zigzag]
Assume the same setting as in EXAMPLE~\ref{eg:ns}.
At the initial time $t_0$, $D$=$\overline D$=$\{3,4,6,7,8,5\}$.
$\overline D_{br}$ are all zeros, and $\overline D_{bw}[i]$ are all ones.
During the first period, transactions $T_1$ and $T_2$ are written to $\overline D$, and $D$ has the time-in-point snapshot data of time $t_0$.
For each write, Zigzag sets $\overline D_{br}[i]$=$ 1$, which means the latest version of page $i$ is in $\overline D[i]$.
At the end of $P_1$ (at time $t_1$), the latest data can be tracked by the $\overline D_{br}$ array.
To take the snapshot, we should ensure that transactions $ T_3 $ and $ T_4 $ cannot write to data tracked by $\overline D_{br}$.
Conversely, $ T_3 $ and $ T_4 $ should write according to $\neg \overline D_{br}$.
So, we set $ \overline D_{bw}$=$\neg \overline D_{br}$.
During the second period $P_2$, we can access the snapshot with the help of $\neg \overline D_{bw}$.
\end{exmp}

\subsubsection{Ping-Pong}
Ping-Pong (PP)~\cite{Cao.13} is proposed to completely eliminate the latency spikes.
It leverages one copy $\overline D_u$ to collect updates and the other copy $\overline D_d$ to record the incremental snapshot.
During each period, the client thread reads from $D$ and writes to both $D$ and $\overline D_u$.
The snapshotter thread can asynchronously access the incremental snapshot $\overline D_d$.
At the end of each period, all the updated data for constructing the upcoming incremental snapshot are held in $\overline D_u$.
PP attains an immediate swap by exchanging the pointers $\overline D_u$ and $\overline D_d$.

\begin{figure}[htb]
	\centering
	\includegraphics[width=0.48\textwidth]{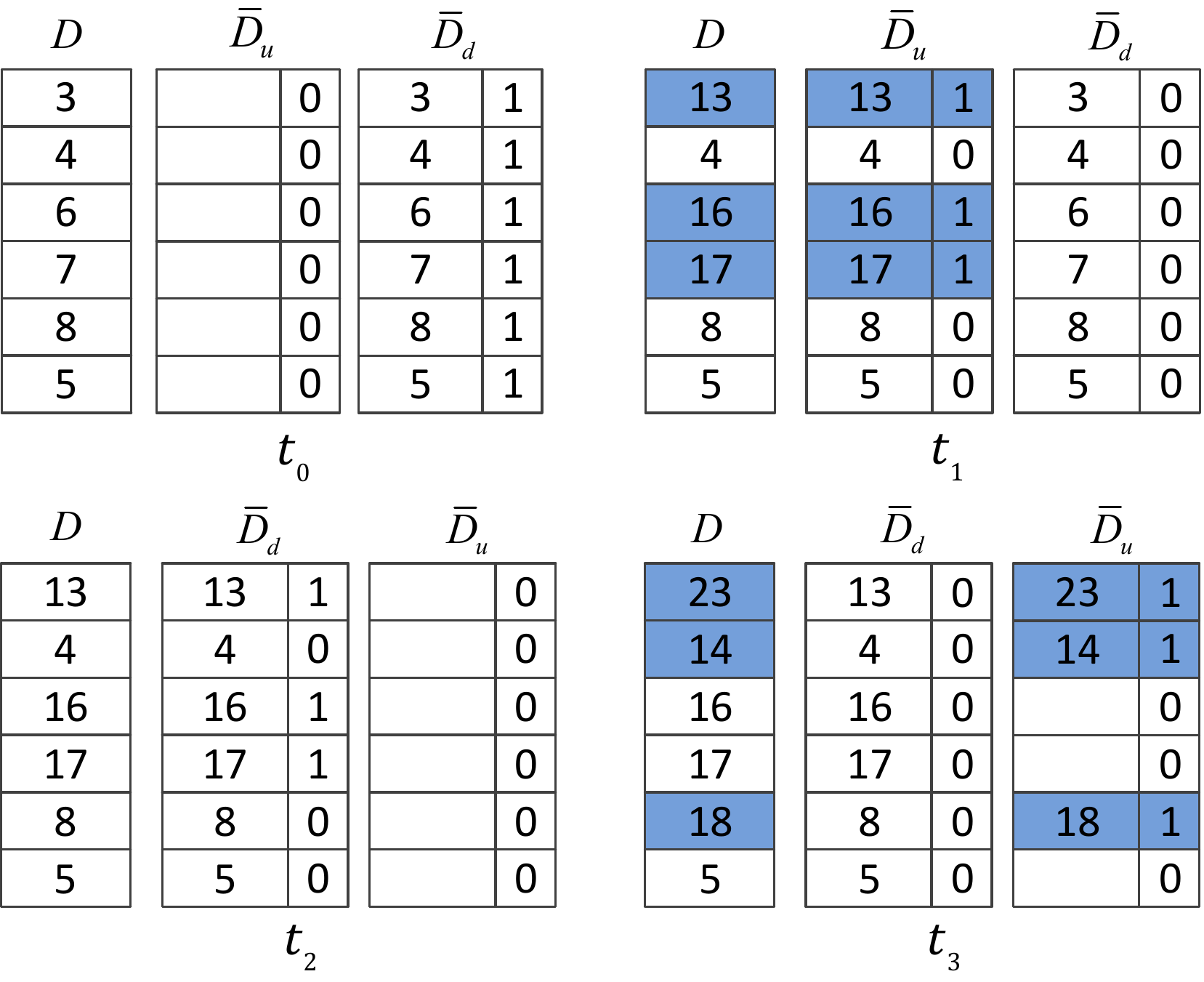}\\
	\caption{Running example for Ping-Pong (PP)}
	\label{fig:pp}
\end{figure}

\begin{figure*}[!htb]
	\centering
	\subfigure[$ t_0 $, $D$ and $\overline D$ with identical dataset]{
		\label{fig:2_a}
		\begin{minipage}[t]{0.22\textwidth}
			\centering
			\includegraphics[width=\textwidth]{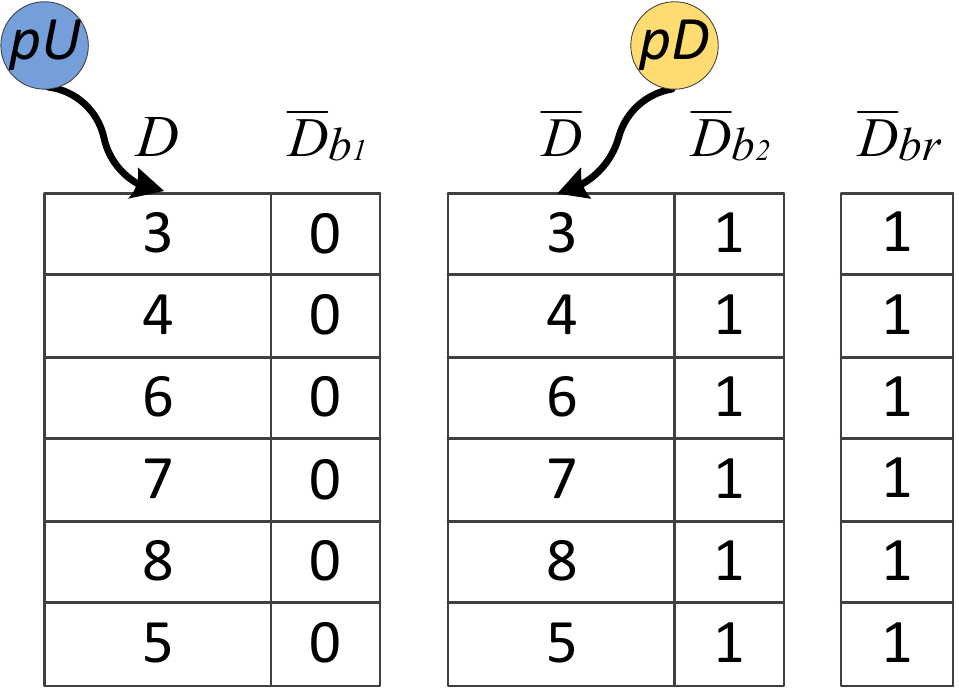}
		\end{minipage}
	}
	\hspace{0.1cm}
	\subfigure[$ t_1 $, the client thread performs updates]{
		\label{fig:2_b}
		\begin{minipage}[t]{0.22\textwidth}
			\centering
			\includegraphics[width=\textwidth]{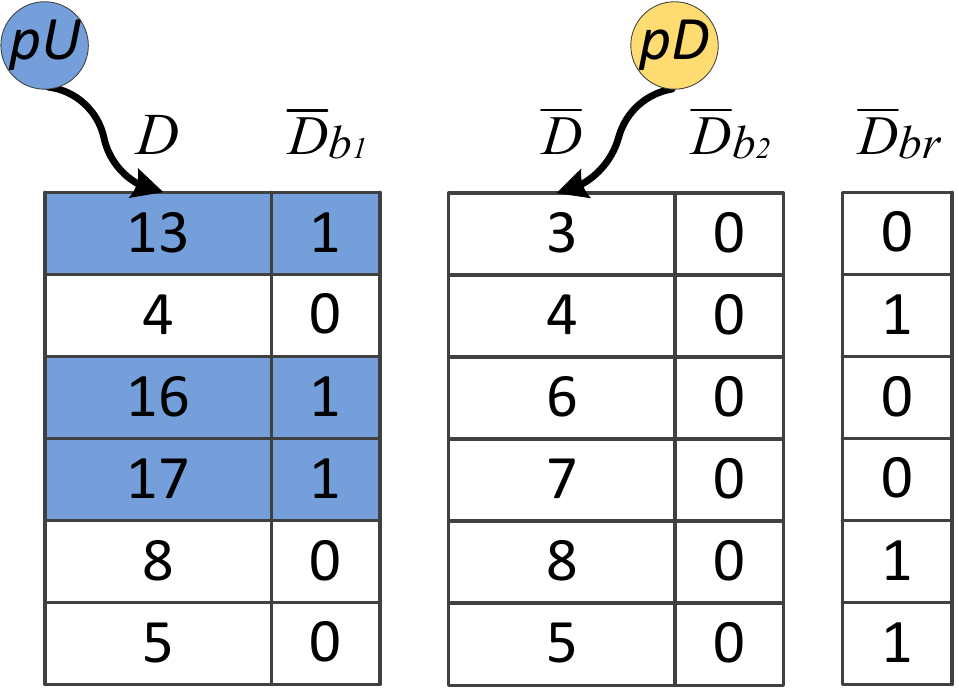}
		\end{minipage}
	}
	\hspace{0.1cm}
	\subfigure[$ t_2 $, the state after pointers swapping]{
		\label{fig:2_c}
		\begin{minipage}[t]{0.22\textwidth}
			\centering
			\includegraphics[width=\textwidth]{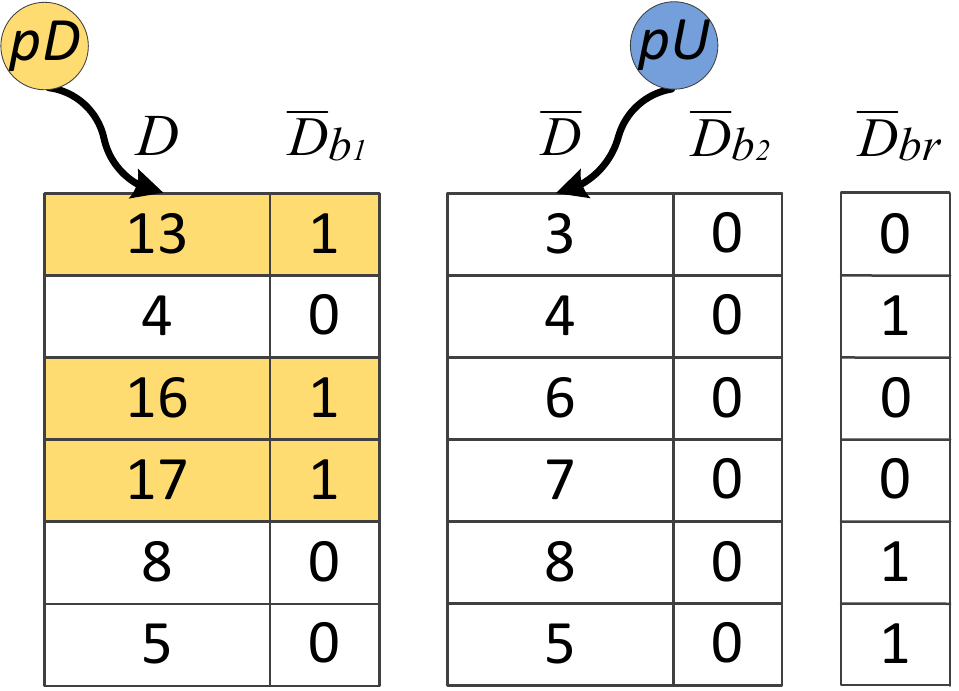}
		\end{minipage}
	}
	\hspace{0.1cm}
	\subfigure[$ t_3 $, the state after backing up the incremental updated data]{
		\label{fig:2_d}
		\begin{minipage}[t]{0.22\textwidth}
			\centering
			\includegraphics[width=\textwidth]{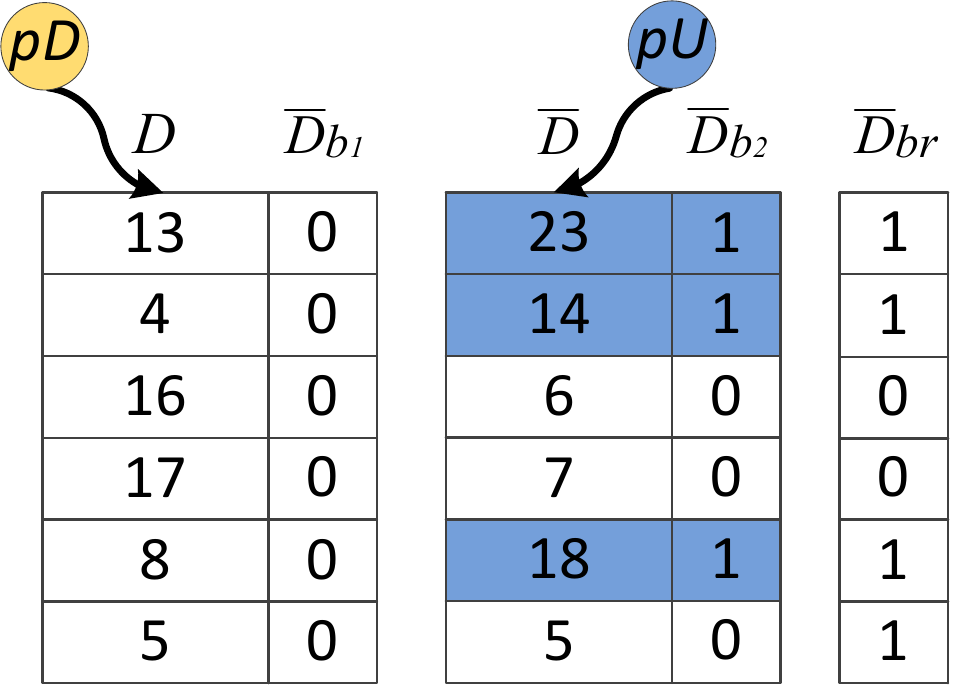}
		\end{minipage}
	}
	\caption{Running example for \LL (\ll)}
	\label{fig:alg_2}
\end{figure*}

\begin{figure*}[!htb]
	\centering
	\subfigure[$ t_0 $, $D$ and $\overline D$ with identical dataset]{
		\label{fig:mk_a}
		\begin{minipage}[t]{0.22\textwidth}
			\centering
			\includegraphics[width=\textwidth]{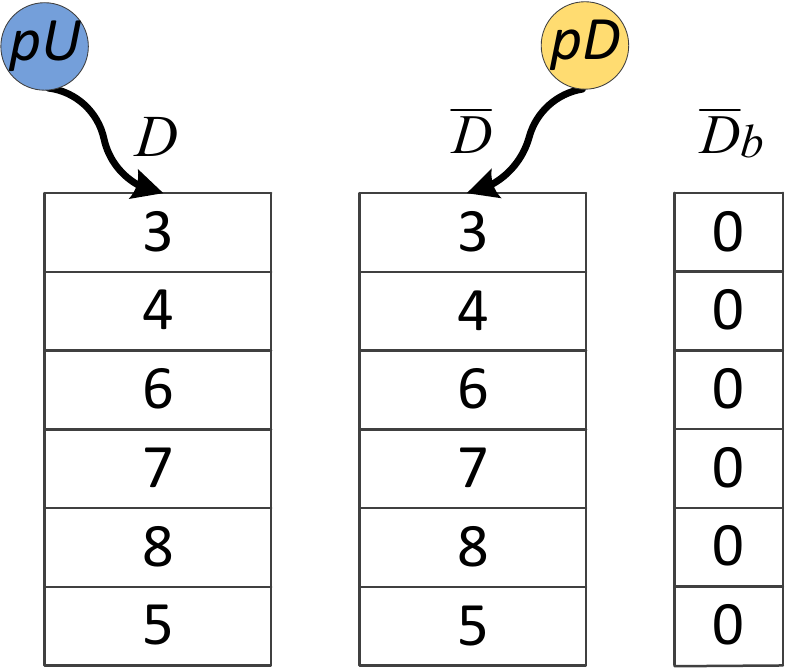}
		\end{minipage}
	}
	\hspace{0.1cm}
	\subfigure[$ t_1 $, the client thread performs updates]{
		\label{fig:mk_b}
		\begin{minipage}[t]{0.22\textwidth}
			\centering
			\includegraphics[width=\textwidth]{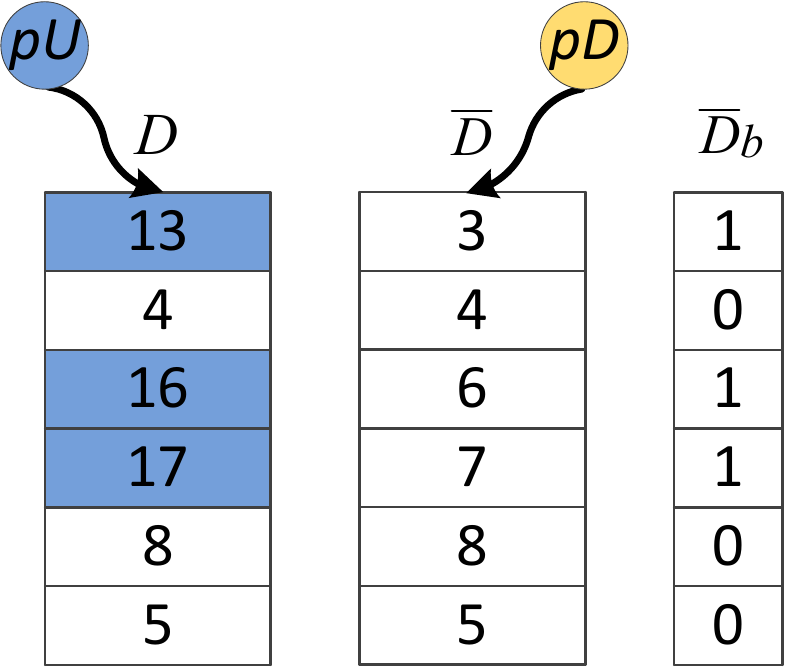}
		\end{minipage}
	}
	\hspace{0.1cm}
	\subfigure[$ t_2 $, the state after pointers swapping]{
		\label{fig:mk_c}
		\begin{minipage}[t]{0.22\textwidth}
			\centering
			\includegraphics[width=\textwidth]{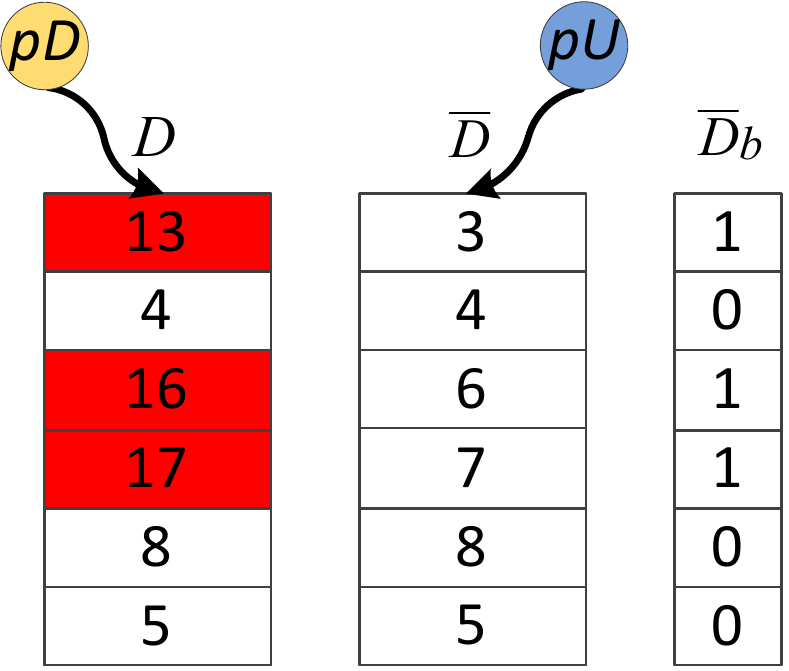}
		\end{minipage}
	}
	\hspace{0.1cm}
	\subfigure[$ t_3 $, the state after backing up the full snapshot]{
		\label{fig:mk_d}
		\begin{minipage}[t]{0.22\textwidth}
			\centering
			\includegraphics[width=\textwidth]{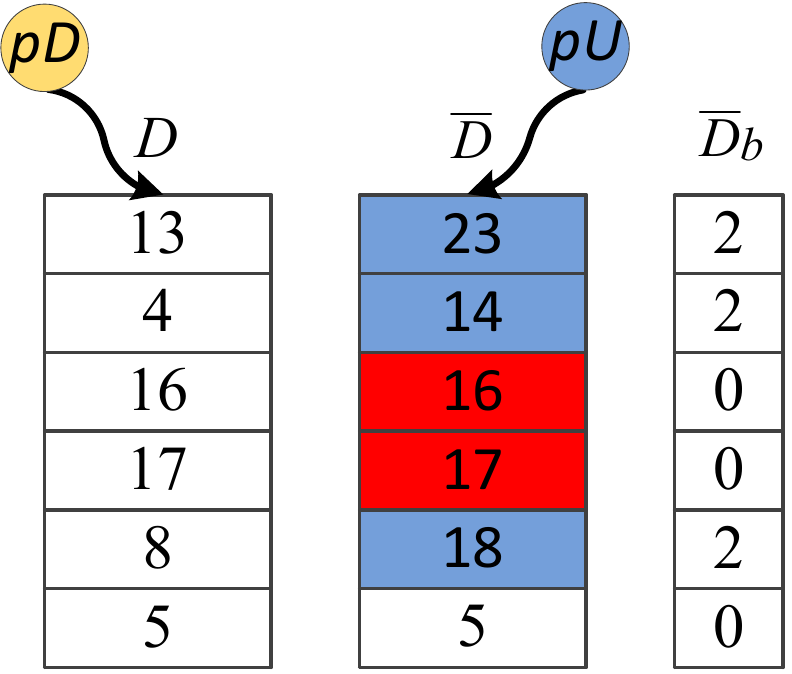}
		\end{minipage}
	}
	\label{fig:alg_mk}
	\caption{Running example for \MK (\mk)}
\end{figure*}

\begin{exmp}[Ping-Pong]
Assume the same setting as EXAMPLE~\ref{eg:ns}.
At time $t_0$, $D$=$\overline D_d$=$\{3,4,6,7,8,5\}$.
During $P_1$, we execute $T_1$ and $T_2$ to $D$ and $\overline D_u$.
Meanwhile, $\overline D_d$ holds the snapshot of time $t_0$.
At the end of $P_1$ (at time $t_1$), the data in $\overline D_u$ hold the incremental data regarding the updated data during $P_1$.
In the taken phase ($t_1 \rightarrow t_2$), we exchange $\overline D_u$ and $\overline D_d$ to freeze the snapshot data.
During $P_2$, we can write data to $\overline D_u$ and access the incremental snapshot in $\overline D_d$.
\end{exmp}

\subsection{Improved Snapshot Algorithms}
Previous snapshot algorithms mainly trade off between latency and throughput.
Fork, a COU variant prevailing in industrial IMDBs, has low latency and high throughput, but the method itself still suffers from high time complexity.
To simultaneously achieve low latency, high throughput, small time complexity and zero latency spikes, we propose two lightweight improvements, \LL and \MK, over existing snapshot algorithms.

\subsubsection{\LL}\label{SubSec:LL}
One intuitive improvement over the above snapshot algorithms is the combination of Zigzag (bit array marking) and Ping-Pong (pointers swapping) to avoid latency spikes while at the same time retaining a small memory footprint.
We call this improvement \LL (\ll).
It maintains dataset $D$ and a shadow copy $\overline D$, which are accessed by pointers ``$pU$'' and ``$pD$'', respectively, as in Ping-Pong.
$D$ and $\overline D$ are accompanied by bit arrays $\overline D_{b1}$ and $\overline D_{b2}$, where $\overline D_{b1}[i]$ and $\overline D_{b2}[i]$ indicate whether the page in $D[i]$ has been updated during the current period.
\LL utilizes these bit arrays to record the incremental data updates in the current period.
Pointer swapping happens at the end of the period.

An additional bit array $\overline D_{br}$ is set up to indicate the locations (either in $D$ or in $\overline D$) from which the client thread can read the latest pages.
A zero for the bit $\overline D_{br}[i]$ indicates that the latest data locate in $D[i]$, and a value of one indicates they are located in $\overline D[i]$.
The following example illustrates how \LL works during two successive snapshots.

\begin{algorithm}[htbp]
	\caption{ \LL}
	\label{alg:LL}

	\begin{algorithmic}
		\Require ~~\\
		DataSet $D, \overline D \leftarrow$ \emph{initial data source}\\
		DataSet $*pU$, $*pD$\\
		BitArray $\overline D_{b1} \leftarrow \{0,0,...,0\}$\\
		BitArray $\overline D_{b2} \leftarrow \{1,1,...,1\}$\\
		BitArray $*pU_b$, $*pD_b$\\
		BitArray $\overline D_{br} \leftarrow \{1,1,...,1\}$\\		
		$D \leftarrow pU$, $\overline D_{b1} \leftarrow pU_b$,
		$\overline D \leftarrow pD$, $\overline D_{b2} \leftarrow pD_b$\\
		$PageNum \leftarrow |D|$
	\end{algorithmic}
%	\rule{8cm}{0.2mm}
	\begin{algorithmic}[1]
		\Function {Client::Write}{$index,newValue$}
		\State $pU_b[index] \leftarrow 1$
		\State $pU[index] \leftarrow newValue$
		\State $\overline D_{br}[index] \leftarrow (pU == \&D)?0:1$
		\EndFunction		
	\end{algorithmic}

	\begin{algorithmic}[1]
		\Function {Client::Read}{$index$}			
			\State \Return $(\overline D_{br}[index] == 0)?D[index]:\overline D[index]$
		\EndFunction
	\end{algorithmic}

	\rule{8cm}{0.2mm}
	
	\begin{algorithmic}[1]
		\Function {Snapshotter::Trigger}{}
		\If{previous snapshot done}		
		\State $ TakeSnapshot() $		
		\State $ TraverseSnapshot() $
		\EndIf
		\EndFunction
	\end{algorithmic}
	
	\begin{algorithmic}[1]
		\Function {Snapshotter::TakeSnapshot}{}
		\State \textbf{lock} Client
		\State \textbf{swap}($pU , pD$)
		\State \textbf{swap}($pU_b , pD_b$)
		\State \textbf{unlock} Client
		\EndFunction
	\end{algorithmic}
	
	\begin{algorithmic}[1]
		\Function {Snapshotter::TraverseSnapshot}{}
		\For{$i=1$ to $PageNum$}
		\If{$pD_b[i] = 1$}
		\State $pD_b[i] \leftarrow 0$
		\State \textbf{write} $pD[i]$
		\Else
		\State \textbf{copy-from} \textit{last snapshot}
		\EndIf
		\EndFor
		\EndFunction
	\end{algorithmic}
\end{algorithm} 

\begin{exmp}[\LL]
As shown in \figref{fig:2_a}, assume that at time $t_0$, $D$=$\overline D$=$\{3,4,6,7,8,5\}$.
$\overline D_{b1}$ and $\overline D_{b2}$ are initialized with zeros and ones, respectively.
$\overline D_{br}$ is initialized with ones.
During $P_1$, when an update occurs on page $i$, $\overline D_{b1}[i]$ is set to 1, and $\overline D_{br}[i]$ is set to 0.
$\overline D$ will be kept away from the client thread, so that it can be accessed by the snapshotter thread in the lock-free manner.
At the same time, once a page $j$ in the dataset $\overline D$ has been accessed, the $j$th position in $\overline D_{b2}$ is reset to 0.
At the end of this period, all bits in $\overline D_{b2}$ are reset to zeros.
\figref{fig:2_b} shows the changes to the memory pages at the end of period $P_1$.
The updated pages are marked in blue shadow.
Next, in the snapshot taken phase, the pointers of $pU$ and $pD$ between $D$ and $\overline D$ are swapped as in \figref{fig:2_c}.
Then, in the access phase, the snapshotter thread begins to access the incremental snapshot data from $D$.
Only those pages pointed by $pD$ where the corresponding bits are set to zeros are included in the snapshot.
In our example, $D[0]$, $D[2]$, and $D[3]$ (marked in yellow shadow \figref{fig:2_c}) are accessed.
During this time, the client thread resumes executing the transactions.
The state at the end of $P_2$ is shown in \figref{fig:2_d}.
%All bits in $\overline D_{b2}$ are set to be zeros.
\end{exmp}

Algorithm \ref{alg:LL} describes the main idea of \LL.
%It implements all five functions defined in the model.
%To make the pseudo code clear, the pointers swapping includes not only that between $pU$ and $pD$, but also that between two pointers $pU_b$ and $pD_b$ over $\overline D_{b1}$ and $\overline D_{b2}$ as shown in Line 2 and Line 3 of Snapshotter::PrepareForCheckpoint().

\subsubsection{\MK}\label{SubSec:MK}
Although pointer swapping (in Ping-Pong and \LL) eliminates latency spikes, it is only applicable for incremental snapshots.
To enable full snapshots with the pointer swapping technique, we propose another improvement called \MK (\mk).
The idea is to copy the out-of-date data from $pD$ to $pU$.
Consequently, the data pointed by $pU$ will always be the latest at the end of each period, \ie $ pD $ holds the full snapshot data after pointer swapping.

\begin{table*}[!htbp]
	\centering
	\caption{Comparison of algorithms in different metrics; ``(*)" represents the drawback}
	\label{tlb:cmp}
	\begin{tabular}{|c|c|c|c|c|c|c|}
		\hline
		{\bfseries Algorithms}  & {\bfseries\makecell{Average \\Latency }} & {\bfseries\makecell{Latency\\Spike}} & {\bfseries\makecell{Snapshot Time \\ Complexity}}  & {\bfseries\makecell{Max \\Throughput}} & {\bfseries\makecell{Is Full \\Snapshot}} & {\bfseries\makecell{Max Memory\\ Footprint}}\\
		\hline
		Naive Snapshot \cite{bronevetsky2006recent,schroeder2007understanding}     & low              &  (*) high  & (*) O(n)  & low     &  yes  & $2\times$              \\
		\hline
		Copy-on-Update \cite{Salles.12,liedes2006siren,Cao2013Fault}     & (*) high   &  (*) middle & (*) O(n)  & middle    & yes    & $2\times$     \\
		\hline
		Fork~\cite{fork-wiki}    & low  & (*) middle   & (*) O(n)  & high    & yes    & $2\times$     \\
		\hline
		Zigzag \cite{Cao.13}      & middle  & (*) middle &(*)  O(n)  & middle    & yes    & $2\times$     \\
		\hline
		Ping-Pong \cite{Cao.13}     & (*) high     & almost none    & O(1)   & low       &  no  & (*) $3\times$   \\
		\hline
		\LL     & low    & almost none    & O(1)    & high       &   no  & $2\times$  \\
		\hline
		\MK    & low    & almost none   & O(1)    & high       &  yes & $2\times$  \\
		\hline
	\end{tabular}
\end{table*}

To support piggyback copies, the \MK algorithm leverages two techniques.
\textit{(i)} \MK maintains a two-bit array $\overline D_{b}$.
The value of $\overline D_{b}[i]$ is one of three states from $\{0, 1, 2\}$, which indicates from which dataset the client thread should read.
When $\overline D_{b}[i]$=$0$, the client thread can read page $i$ from either array because it means that $D[i]$=$\overline D[i]$.
When $\overline D_{b}[i]$=$1$, the client thread should read page $i$ from $D[i]$.
When $\overline D_{b}[i]$=$2$, the client thread should read page $i$ from $\overline D[i]$.
\textit{(ii)} \MK defines another function \textbf{Snapshotter::WriteToOnline()} which is called in \textbf{Snapshotter::Trigger()} as in Algorithm \ref{alg:MK}.
\textbf{Snapshotter::WriteToOnline()} ensures the data pointed by $ pU $ will always be the latest at the end of each period,
so that \textbf{Snapshotter::TraverseSnapshot()} can access the full snapshot in $ pD $.

\begin{algorithm}[htbp]
	\caption{ \MK}
	\label{alg:MK}

	\begin{algorithmic}
		\Require ~~\\
		DataSet $D, \overline D  \leftarrow$ \emph{initial data source}\\
		DataSet $*pU$, $*pD$\\
		$D \leftarrow pU$, $\overline D \leftarrow pD$\\
		FlagArray $\overline D_{b} \leftarrow \{0,0,...,0\}$ \\
		$PageNum \leftarrow |D|$
	\end{algorithmic}
%	\rule{8cm}{0.2mm}

	\begin{algorithmic}[1]
		\Function {Client::Write}{$index,newValue$}
		\State $pU[index] \leftarrow newValue$
		\State $\overline D_{b}[index] \leftarrow (*pU \neq D)?2:1$
		\EndFunction
	\end{algorithmic}

	\begin{algorithmic}[1]
		\Function {Client::Read}{$index$}
		\State \Return $(\overline D_{b}[index] \neq 2)?D[index]:\overline D[index]$
		\EndFunction
	\end{algorithmic}

	\rule{8cm}{0.2mm}

	\begin{algorithmic}[1]
		\Function {Snapshotter::Trigger}{}
		\If{previous snapshot done}		
		\State $ TakeSnapshot() $
		\State $ WriteToOnline() $		
		\State $ TraverseSnapshot() $
		\EndIf
		\EndFunction
	\end{algorithmic}

	\begin{algorithmic}[1]
		\Function {Snapshotter::TakeSnapshot}{}
		\State \textbf{lock} Client
		\State \textbf{swap}($pU , pD$)
		\State \textbf{unlock} Client
		\EndFunction
	\end{algorithmic}

	\begin{algorithmic}[1]
		\Function {Snapshotter::WriteToOnline}{}
		\State $bit = (pD == \&D)?1:2$
		\For{$k=1$ to $PageNum$}
		\If{$\overline D_{b}[k]=bit$}
		\State $\overline D_{b}[k]=0$
		\State $pU[k] \leftarrow pD[k]$
		\EndIf
		\EndFor
		\EndFunction
	\end{algorithmic}

	\begin{algorithmic}[1]
		\Function {Snapshotter::TraverseSnapshot}{}
		\For{$k=1$ to $PageNum$}
		\State \textbf{Dump-All} $pD[k]$
		\EndFor
		\EndFunction
	\end{algorithmic}
	
\end{algorithm} 

\begin{exmp}[\MK]
Initially, $pU$ and $pD$ are pointed to $D$ and $\overline D$, respectively.
The bit array $\overline D_{b}$ is set to zeros as shown in \figref{fig:mk_a}.
\figref{fig:mk_b} shows the situation at time $t_1$.
The client thread updates pages $D[0]$, $D[2]$, and $D[3]$ (blue shadow) during the first period.
The corresponding two-bit elements in $\overline D_{b}$ are then set to ones by the client thread at the same time.
This ensures that the client thread always reads the latest data based on the information in $\overline D_{b}$.
Concurrently, $\overline D$ has the full snapshot data of time $t_0$.
%Since there is no newer data in corresponding pages of $\overline D$ than those of $D$, we need not piggyback copy from $\overline D$ to $D$.

At the beginning of $P_2$, pointers $pU$ and $pD$ are exchanged.
A full snapshot about time $t_1$ is held in this copy in $D$ and can be accessed.
Meanwhile, $\overline D$ can be updated by the client thread.
Note that there may be dirty pages in $\overline D$ in the $P_3$ period.
For instance, $\overline D[0]$, $\overline D[2]$ and $\overline D[3]$ (red shadow) are older pages (\figref{fig:mk_c}).
To avoid dirty pages, \MK performs a piggyback copy of these pages from $D$ to $\overline D$ in this period together with the client's normal updates on pages $\overline D[0]$, $\overline D[1]$ and $\overline D[4]$ (blue shadow).
Hence, at the end of $P_2$, all the pages in $\overline D$ are updated to the latest state as shown in \figref{fig:mk_d}.
%State bit array $\overline D_b$ are set accordingly to indicate from which dataset can the client thread read the latest data.
\end{exmp}

\subsection{Comparison of Snapshot Algorithms}\label{sec:comp}
\tabref{tlb:cmp} compares the advantages and drawbacks of the snapshot algorithms.
Although fork is a variant of COU, we list it separately since it is the standard method in many industrial IMDBs.
In theory, \MK, our modification over Zigzag and Ping-Pong, outperforms the rest in all metrics.

Note that the $2\times$ memory consumptions of \ll and \mk are only for the abstract array model (static memory allocation).
Their memory footprints can be further reduced in the production environment thanks to the dynamic memory allocation technique (see \secref{Sec:SystemValidations}).
%The main reason is that in the array model, the memory are pre-allocated in $2\times$ or $3\times$, but dynamic-allocate memory is suitable for real systems.
%\TODO{more description}
\section{Virtual Snapshot}\label{sec:discuss}
This section discusses the recent work~\cite{ren2016low-overhead} in designing virtual snapshot algorithms that are independent of a physically consistent state.
We also modify our \LL and \MK algorithms to meet this new requirement.
%and opensource the implementation on GitHub \footnote{\url{https://github.com/bombehub/VirtualSnapshot}}.
\subsection{Physical snapshot algorithms with Physically Consistent State}
The above snapshot algorithms from academia and industry rely on a physically consistent state. That is, the in-memory data must remain consistent at a point in time once the trigger function is invoked. Such a situation has been discussed frequently in applications, such as frequent consistent application~\cite{Salles.12,Cao.13}, actor-oriented database systems~\cite{actor-database}, and partition-based single thread running database,\ie H-Store~\cite{kallman2008h-store:}, Redis, Hyper~\cite{kemper2011hyper}, etc. 

However, for a broader application situation (e.g., concurrent transaction based database), to maintain such a physically consistent state, system quiescing is inevitable until all active transactions have been committed. This is the cause of latency spikes~\cite{ren2016low-overhead}.

\subsection{Virtual Snapshot Algorithms without Physically Consistent State}
\subsubsection{CALC}
To avoid blocking transactions during the trigger, one recent pioneering work called CALC~\cite{ren2016low-overhead} proposes the concept of virtual consistent snapshot,
for which snapshot is not captured at the point in trigger time but delayed until the end of all active transactions.
CALC is a concurrent variant of COU.
In CALC, each cycle (period) is divided into 5 phases.
Similar to COU, CALC maintains two copies of data $D$ and $\overline D$, as well as a bit array $\overline D_b$.
CALC can obtain a virtual consistent view of the snapshot data by carefully performing COU during specific phases.
We interpret the idea through the following example.

\begin{figure}[htb]
	\centering
	\includegraphics[width=0.48\textwidth]{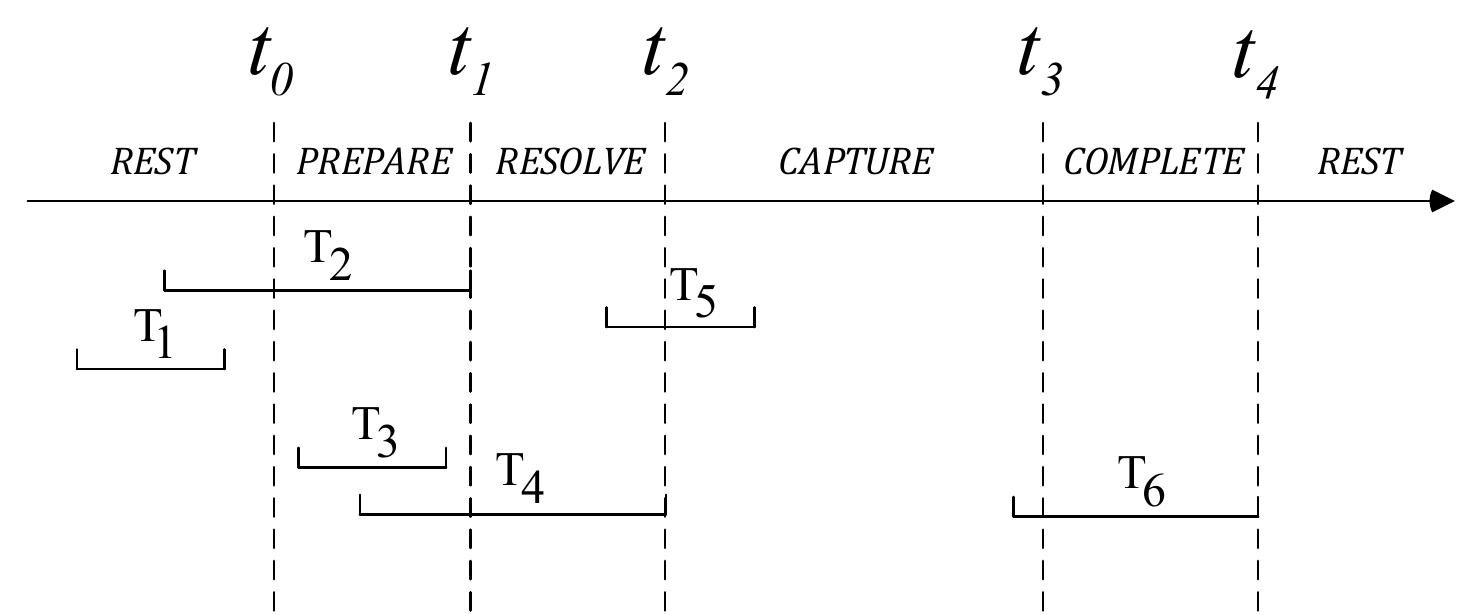}\\
	\caption{Running example for CALC}
	\label{fig:calc}
\end{figure}

\begin{exmp}[CALC]
	As shown in \figref{fig:calc}, the trigger is invoked at time $t_0$.
	The time before $t_0$ is the rest phase.
	At time $t_1$, all the transactions started in the rest phase are committed.
	The time interval $t_0 \rightarrow t_1$ is called the prepare phase.
	At time $t_2$, all the transactions started in the prepare phase are committed, and the corresponding time interval $t_0 \rightarrow t_1$ is labeled the resolve phase.
	The snapshot is taken during $t_2$ to $t_3$, which is called the capture phase.
	At time $t_4$, all the transactions started in the capture phase are committed.
	The time interval $t_3 \rightarrow t_4$ is labeled as the complete phase.
	
	For transactions ($T_1$, $T_2$) started during the rest or the complete phase, CALC only needs to update $D$.
	For transactions ($T_3$, $T_4$, $T_5$, $T_6$) started during the prepare, the resolve or the capture phase, CALC performs the COU strategy.
	Finally, a virtual consistent view snapshot is generated at time point $t_1$.
	The virtual consistent view of the snapshot data should contain $T_1$, $T_2$, $T_3$, and we can start accessing the view of data after $t_2$.
\end{exmp}

% Although our work focuses on checkpointing algorithms for the serial transaction processing as stated in Section \ref{Sec:Introduction},
% both HG and PB can be easily extended to execute in  multi-client scenarios.
% without the limitation of the tick model in Section \ref{Subsec:DefinitionModel}.
% Just like CALC, no transaction blocking happens during the prepare phase of the concurrent version of HG or PB.
% 不同的地方在与updater线程从一个变成了多个
% 我们仍然让这多个updater线程访问的数据和dumper分别访问不同的数据。
% 因此dumper线程可以无锁的后台备份。
% 对于多个并发的updater线程，我们并不关心是如何处理的。可以采用严格两阶段锁，或者是时间戳等方法。
% 事务从开始到提交，需要一直在访问事务开始时刻的那一份数据
% 也就是说：指针交换后，还未提交的活动事务并不受影响。指针交换后才开始的事务，会操作交换后的pU所在的数据.

\subsubsection{vHG and vPB}
Although our improved snapshot algorithms \LL and \MK are primarily designed to be dependent on a physically consistent state, we find such a dependency can be easily eliminated.
We call the new versions of \LL and \MK vHG and vPB, respectively.

We describe the main idea of vHG as follows.
The trick here is that once the trigger function is invoked, the pointers are swapped immediately.
The new transactions (\ie those started after the trigger) should update the data pointed by $pU$ while the active transactions (\ie those uncommitted when the trigger is invoked) will keep updating the data pointed by $pD$.
In other words, pointer swapping does not influence the writing strategies of active transactions.
The access operation of the snapshotter thread should wait until all active transactions are committed.
Note that vPB shares the similar idea with vHG.

%Similar to of CALC, both HG and PB can be extended to concurrent situation,
%we labeled them vHG and vPB.
%The main difference between the concurrent and serial situation is that
%once a trigger() is invoked and pointers are swapped immediately, the newly coming transactions (a.k.a begin after trigger)  should update the dataset pointed by $pU$ while those active transactions (a.k.a. uncommitted when trigger invoked)  will keep updating the dataset pointed by $pD$.
%The asynchronous access operation of snapshotter thread should wait until all active transactions are committed.

\begin{figure}[htb]
	\centering
	\includegraphics[width=0.48\textwidth]{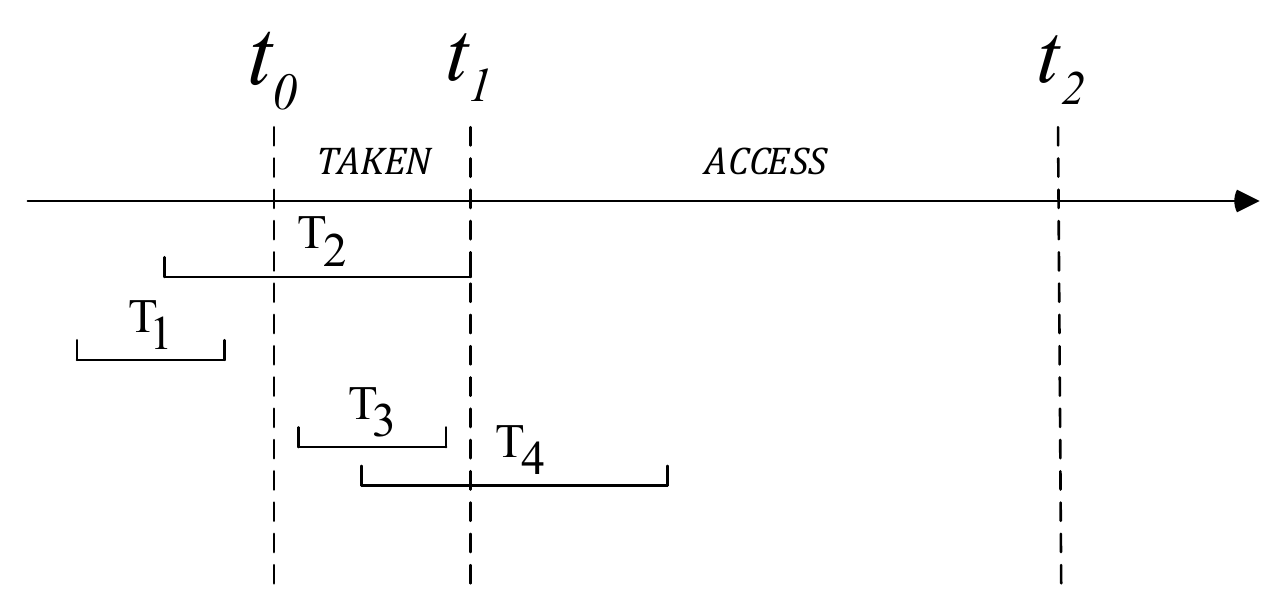}\\
	\caption{Running example for vHG}
	\label{fig:vhg}
\end{figure}

\begin{exmp}[vHG]
	As shown in \figref{fig:vhg}, the structure of vHG is the same as HG.
	The main difference lies in the trigger function.
	At time $t_0$ when the trigger is invoked, the pointers of $pU$ and $pD$ are swapped immediately.
	The transactions started before $t_0$ ($T_1$, $T_2$) are updated to $pU$ (\ie $D$) regardless of the pointer swapping.
	In contrast, the transactions started after $t_0$  ($T_3$, $T_4$) are updated after swapping $pU$ (\ie $\overline D$). % Editor: Please ensure the meaning has been preserved.
	Once $T_1$ and $T_2$ are committed at time $t_1$, the data in $pD$ hold the virtual consistent view of data.
	Then, the snapshotter thread invokes the TraverseSnapshot() function.
\end{exmp}

\begin{algorithm}[htbp]
	\caption{ vHG}
	\label{alg:vhg}
	
	\begin{algorithmic}[1]
		\Function {Client::TransactionExecution}{$txn$}
		\If{$pU$ equals to $D$}		
    		\For{$index, newvalue$ in $txn$}
        		\State $D_b[index]$ = 1
                \State $D[index]$ = newvalue
    		\EndFor
        \Else
            \For{$index, newvalue$ in $txn$}
        		\State $\overline D_b[index]$ = 1
                \State $\overline D[index]$ = newvalue
    		\EndFor
		\EndIf
		\EndFunction
	\end{algorithmic}

	\begin{algorithmic}[1]
		\Function {Snapshotter::Trigger}{}
		\If{previous snapshot done}		
		\State $ TakeSnapshot() $	
        \State $ Detect\_and\_Waiting() $	
		\State $ TraverseSnapshot() $
		\EndIf
		\EndFunction
	\end{algorithmic}

\end{algorithm} 

%\liliang{Algorithm \ref{alg:vhg} describle the detail of vHG algorithms. the main framework is the same as Algorithm \ref{alg:LL}, the difference is when performing updates within a transaction, we should updating date to the same dataset all the lifecycle. and when the trigger is invoked, we should not performe snapshot immidietly, but need to detect and waiting for the actived transactions to finished, at the same time, the new comming transaction can run  normally, as shown in line 4 of  snapshotter::trigger() function.}

Algorithm~\ref{alg:vhg} shows the pseudo code of vHG, which shares the same framework with Algorithm ~\ref{alg:LL}. The difference lies in the fact that data should be updated to the same dataset within a transaction lifecycle, and snapshot should be postponed by detecting and waiting for the end of all active transactions rather than being performed immediately after the trigger is invoked. In this way, incoming transactions cannot be blocked as shown in line 4 of \textbf{Snapshotter::Trigger()} function.

\section{Experimental Studies}
This section comprehensively evaluates the performance of various snapshot algorithms from the previous section.
We first present a thorough benchmark study on latency, throughput and snapshot overhead (\secref{sec:comparison}).
In addition, we briefly evaluate the performance of virtual snapshot algorithms(\secref{SubSec:extend_exp}).
Then, we implement two Redis variants by integrating \ll and \mk, respectively, to study the scalability in real-world IMDB systems (\secref{Sec:SystemValidations}).
%The experiments have 2 part.
%The first part,  we implement all the algorithms within the consistent checkpoint application to let the algorithms be consist with \cite{Cao.13}.
%Our comparisons (Section \ref{sec:comparison}) cover client thread overhead which include latency, maximum throughput and checkpointing overhead of the snapshotter thread.
%We also adapt our HG and PB algorithms to the concurrent processing situation,
%and then compare vHG and vPB throughput with CALC.
%The second part (Section \ref{Sec:SystemValidations}), we implement two redis variants through separately integrate HG and PB with redis,
%form two good scalability prototype: Redis-HG and Redis-PB.
%We conclude this section by summarizing our experimental results (Section \ref{sec:summary}).

\subsection{Infrastructure}\label{SubSec:Settings}
All the experiments are conducted on a server, HP ProLiant DL380p Gen8, which is equipped with two E5-2620 CPUs and 256GB main memory.
% Totally, the server has 24 logical threads because of the Hyper-Threading technique.
% Each core has a private 32KB L1 data cache, 32KB L1 instruction cache, and 256KB L2 cache.
% Moreover, each group of 6 cores on a single sockets share a 15MB L3 cache.
%Total 256GB of DDR3 DRAM is equally divided and attached to each of the CPU. %under the NUMA architecture.
CentOS 6.5 X86\_64 operating system with Linux kernel 2.6.32 and GCC 5.1.0 os installed.
Using the micro-benchmark~\footnote{\url{http://www.cs.cornell.edu/bigreddata/games/recovery_simulator.php}}, the evaluation environment has the following performance parameters:
memory bandwidth = 2.72 GB/s,
memory write startup overhead = 37.38 ns,
lock overhead = 87.19 ns,
and atomic bit check overhead = 0.94 ns.

\subsection{Benchmark Study of Physical Snapshot}\label{sec:comparison}
This trial of experiments evaluates snapshot algorithms with synthetic update-intensive workloads.
\subsubsection{Setups}
We benchmark all snapshot algorithms in checkpoint applications to reveal performance and follow the setups in~\cite{Cao.13}.
The update-intensive client is simulated by a Zipfian \cite{gray1994quickly} distribution random generator.
It generates a stream of update data (in the form of $<page\_index, value>$), which will be consumed by the client thread.
The generator ensures only a small portion of data to be ``hot'', \ie, frequently updated.
Since the Zipfian distribution parameter $\alpha$ has little impact on the experimental results~\cite{Cao.13}, we set $\alpha$ to $2$ by default.
To simulate a heavy updating workload, all the synthetic data are pre-generated and kept in a trace file.
The trace file is loaded into the main memory before performing updates.

To carefully control the update frequency, the client thread runs in a tick-by-tick (a.k.a. time slice) way~\cite{Cao.13}, and we divide each tick into two stages.
One is the \textit{update} stage in which $uf$ times updates should be accomplished.
The length of the update stage is defined as the tick latency (\textit{latency} for short).
Latency will be used as one of the evaluation metrics for performance comparison.
The remaining duration of a tick is regarded as the \textit{idle} stage, which aims to idle the client thread until the start of the next tick to guarantee a consistent tick duration of 100ms.
We can control the update frequency by adjusting the proportion of the two stages.
For example, assume a memory page size of 4KB.
Each page contains 1024 data items, and the data item is 4 bytes in size.
Then, $D$ has 1,000,000 pages, approximately 4000MB, and the data are updated at a rate of 0.128\% per second.
Hence, the update frequency is $\frac{4000MB}{4B}$$\times$0.128\%=1280K per second, \ie $uf$=128K per tick.

The checkpoint interval is defined to be at least 10 seconds.
For each experiment, we monitor 10 successive checkpoints.

\tabref{tlb:parameter} summarizes the parameters of the synthetic workloads.
The initial parameter settings are shown in bold.
Two tunable parameters, \textbf{\textit{dataset size}} and \textbf{\textit{uf}}, can notably affect the performance of the algorithms.
Intuitively, the size of the dataset directly relates to the checkpointing overhead and the time of the snapshotter taken phase.
Furthermore, during each tick interval, we perform accurate times of operations, which represents the intensity of the client workload.
Apparently, the workload has very strong influence on the latency performance.
Hence, we conduct experiments to quantify the impacts of these two parameters in the following.

\begin{table}[!htb]\small
	\centering
	\caption{Parameters of synthetic workloads}
	\label{tlb:parameter}
	\begin{tabular}{|c|c|}
		\hline
		{\bfseries Parameters} &   {\bfseries Setting} \\
		\hline
		Checkpoint count & 10 \\
		\hline
		Data item size &       4B \\
		\hline
		Memory page size &          4KB \\
		\hline
		Tick length &  100ms \\
		\hline
		Update frequency &  \textbf{16k}, 32k, 64k, 128k, 256k \\
		\hline
		Dataset size & \textbf{1000MB}, 2000MB, 4000MB, 8000MB \\
		\hline
	\end{tabular}
\end{table}

\begin{figure*}[!htb]
	\centering
	\begin{minipage}[t]{0.32\textwidth}
		\centering
		\includegraphics[width=\textwidth]{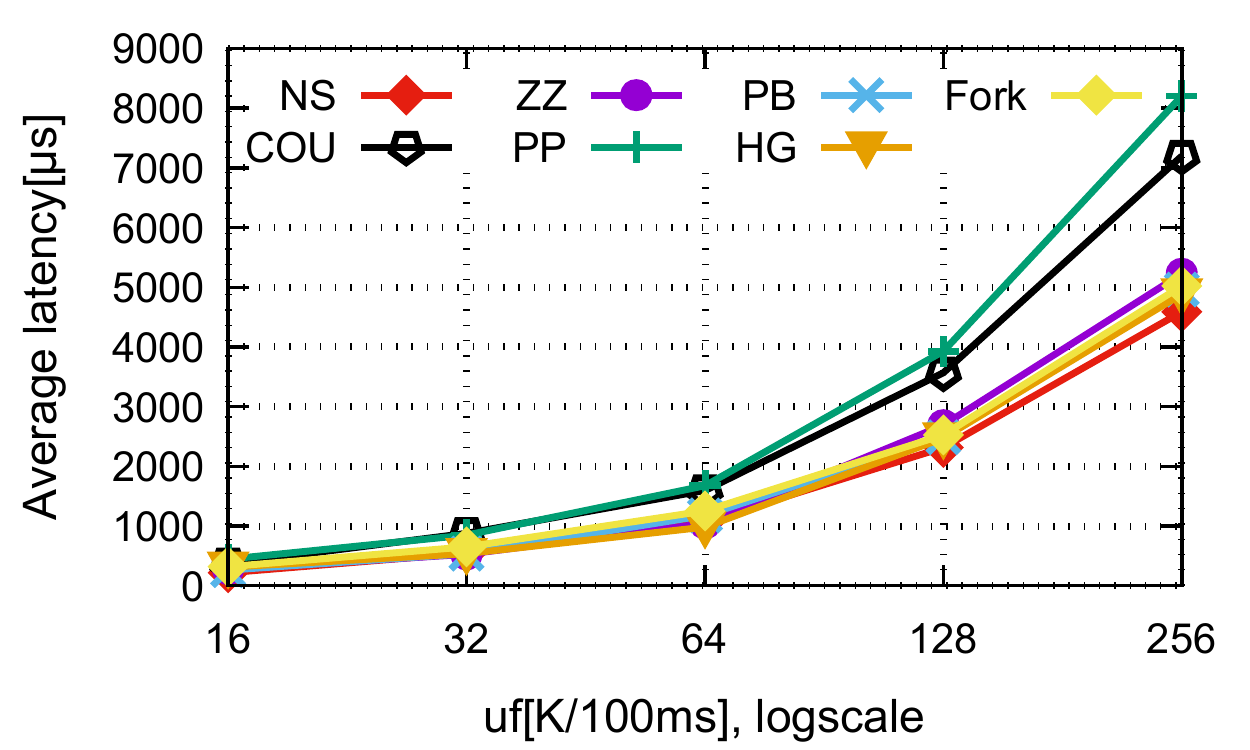}
		\caption{\textbf{\textit{uf}} vs. Average latency}
		\label{fig:uf:latency}
	\end{minipage}
	\hspace{0.1cm}
	\begin{minipage}[t]{0.32\textwidth}
		\centering
		\includegraphics[width=\textwidth]{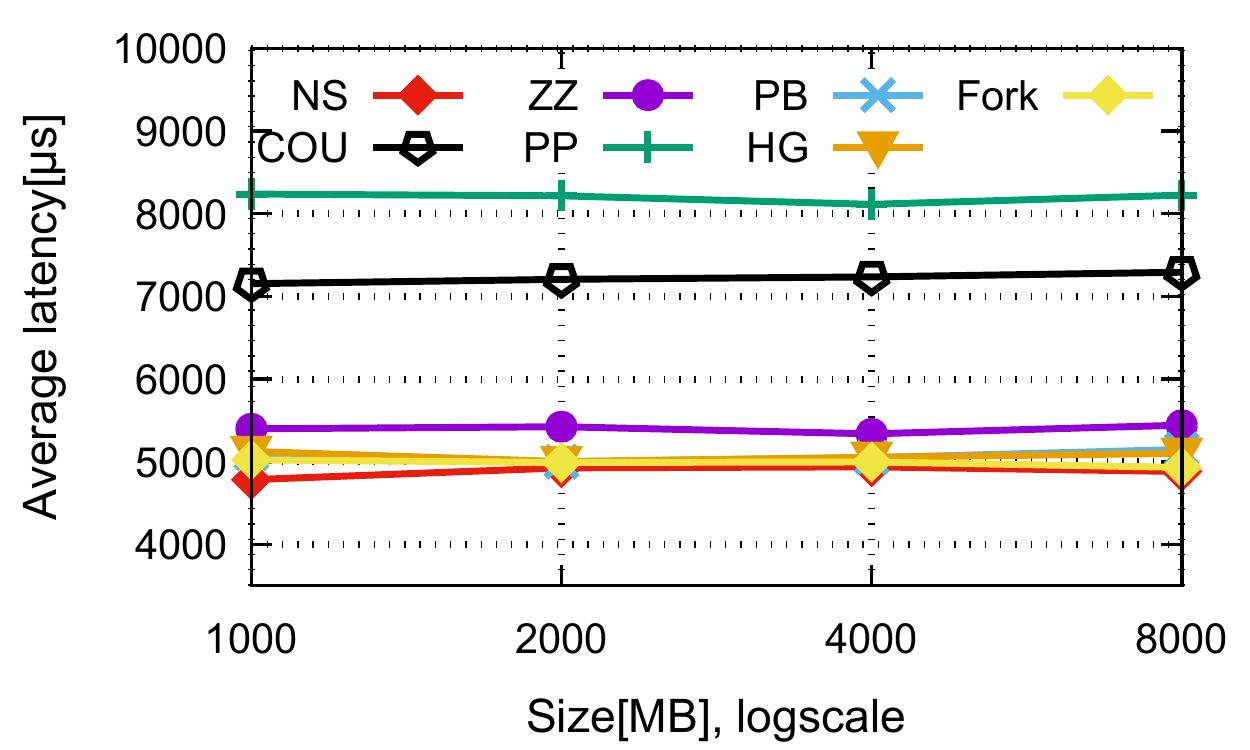}
		\caption{Data size vs. Average latency}
		\label{fig:size:latency}
	\end{minipage}
	\hspace{0.1cm}
	\begin{minipage}[t]{0.32\textwidth}
		\centering
		\includegraphics[width=\textwidth]{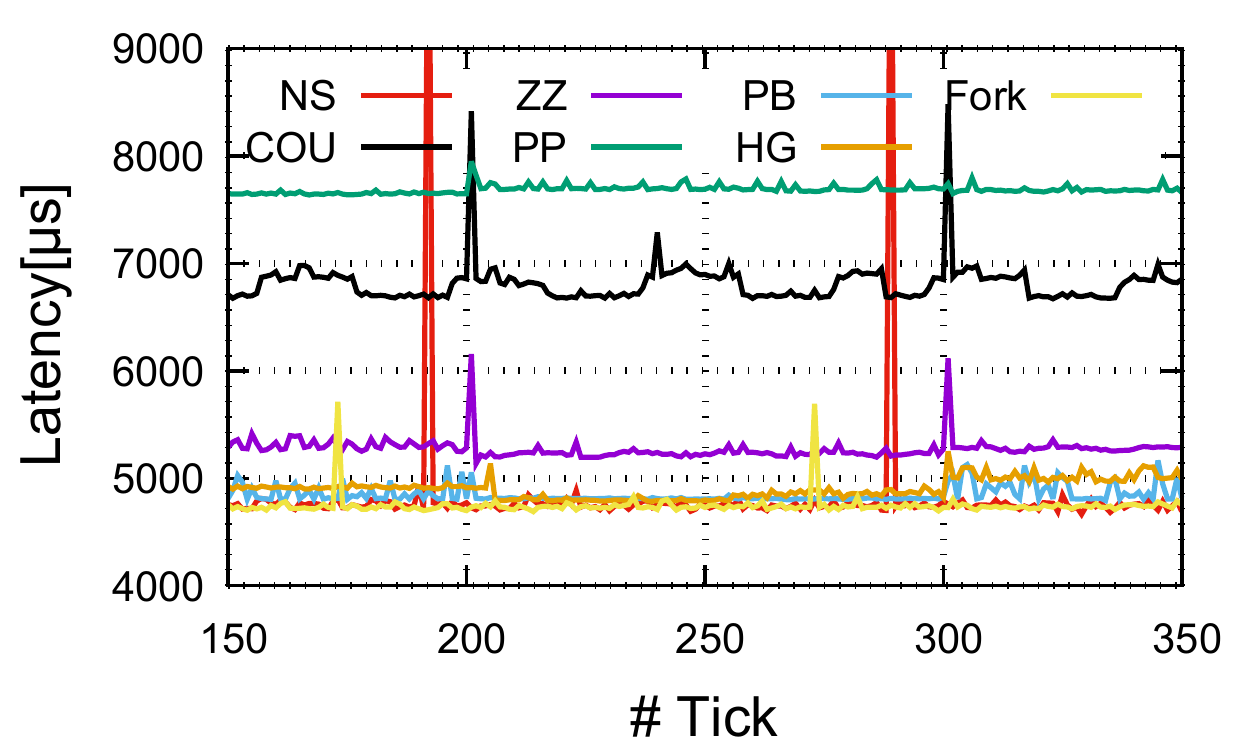}
		\caption{Latency distribution (\textbf{\textit{uf}}=256K)}
		\label{fig:256}
	\end{minipage}
\end{figure*}

\begin{figure*}[!htb]
	\centering
	\begin{minipage}[t]{0.32\textwidth}
		\centering
		\includegraphics[width=\textwidth]{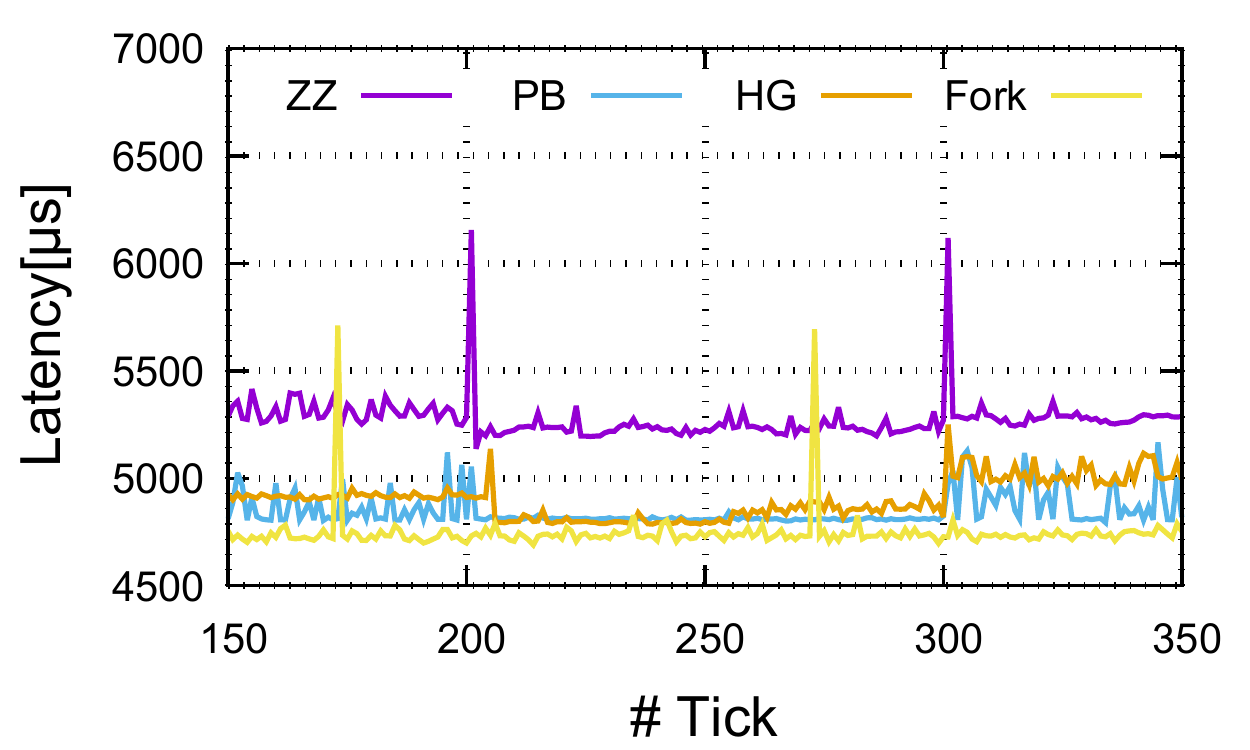}
		\caption{Part of Figure \ref{fig:256} without NS, COU, and PP}
		\label{fig:256-part}
	\end{minipage}
	\hspace{0.1cm}
	\begin{minipage}[t]{0.32\textwidth}
		\centering
		\includegraphics[width=\textwidth]{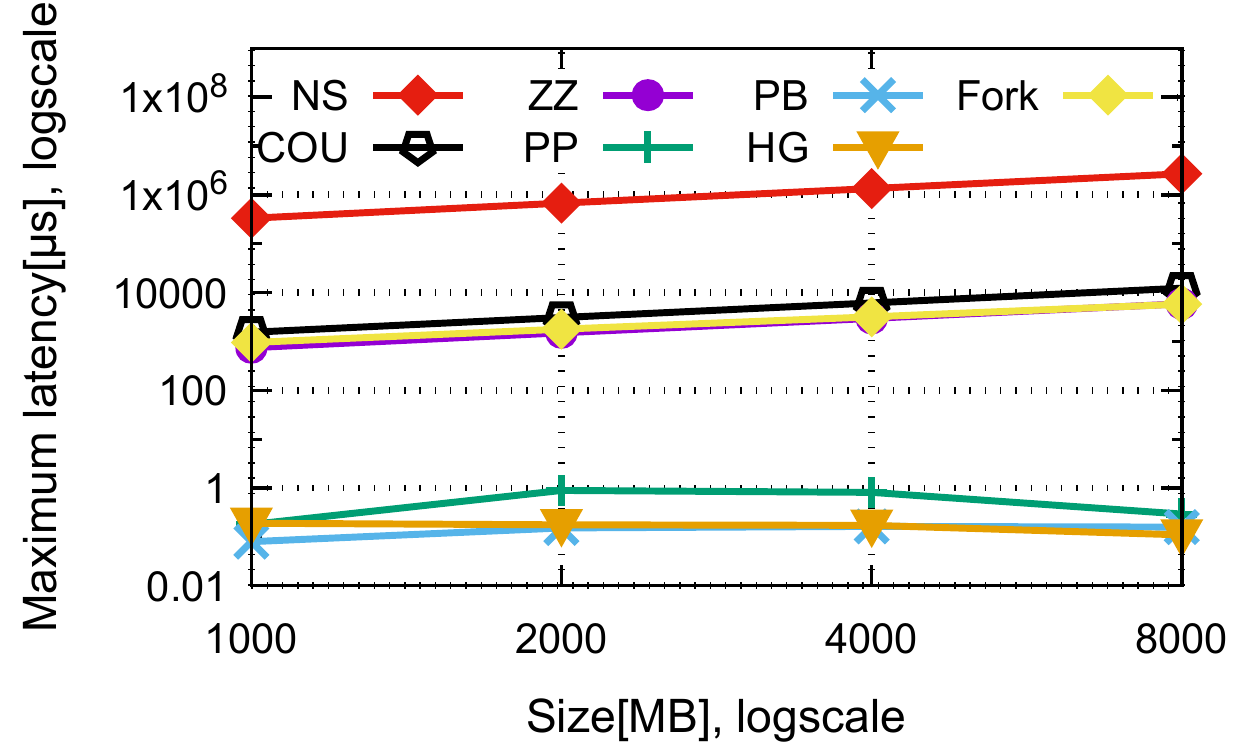}
		\caption{Data size vs. Maximum latency}
		\label{fig:size:prepare}
	\end{minipage}
	\hspace{0.1cm}
	\begin{minipage}[t]{0.32\textwidth}
		\centering
		\includegraphics[width=\textwidth]{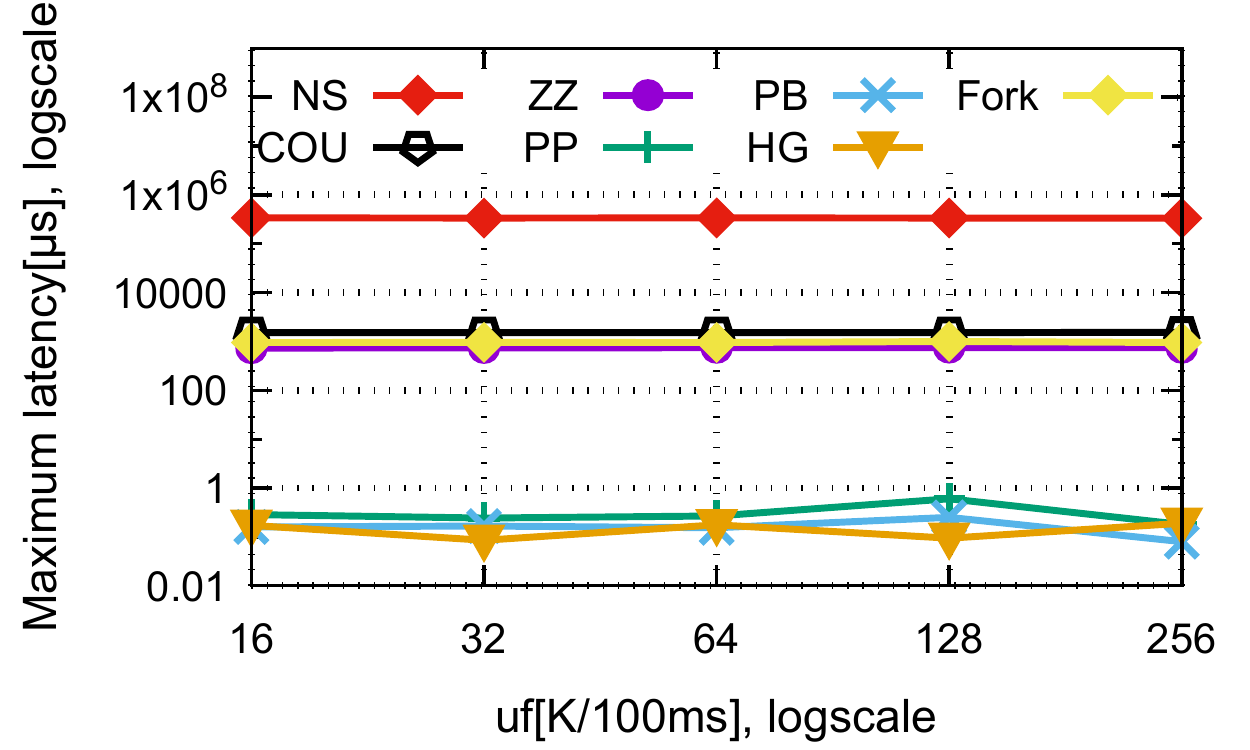}
		\caption{\textbf{\textit{uf}} vs. Maximum latency}
		\label{fig:uf:prepare}
	\end{minipage}
\end{figure*}

\subsubsection{Performance}\label{SubSec:SyntheticLatency}
We mainly evaluate the latency (average, distribution and maximum), maximum throughput and checkpoint overhead of different snapshot algorithms.

\textbf{Average Latency.}
\figref{fig:uf:latency} shows the average latency with the increase of update frequency (16k, 32k, 64k, 128k, and 256k per tick) on a 1000MB dataset.
The average latencies of all algorithms exhibit similar increasing trends.
NS has the shortest average latency because the normal read and update show no interference by additional copy or bit checking operations.
COU has a long latency because there are synchronization locks on pages to be updated between the client and the snapshotter.
Page locking and duplicating increase latency.
Unlike~\cite{Cao.13}, we observe that PP incurs a large latency,
as PP exploits the redundant update mode for the client.
That is, the client thread of PP has to update both $D$ and $\overline D_u$ during each operation.
The experiments in \cite{Cao.13} only consider one of the writes, while we also take the redundant writes into account.
The same result can be found in \cite{ren2016low-overhead}.
The latency of \mk, \ll, Fork and ZZ is relatively small.
In the case of \mk, \ll, and ZZ, they only need an extra bit operation rather than the costly page replication.
The slightly larger latency of ZZ is because each update operation may occur at either dataset ($D$ or $\overline D$), while the other two methods (HG and PB) have an exclusive dataset for updating during a checkpointing period.
The accumulative cost of such interlaced memory addressing cannot be ignored in the case of high updating frequencies.
As for Fork, the OS kernel function \textit{fork}() is called to create a copy of the updating process context including the memory space, which is more efficient than the original COU.
%The straightforward implementation of COU involves much more user mode heavyweight locks.

\figref{fig:size:latency} shows the average latency on datasets of different sizes at a fixed updating frequency of 256K.
We observe that the data size has almost no impact on average latency.

Referring to \figref{fig:uf:latency} and \figref{fig:size:latency}, we conclude that \ll, \mk and Fork exhibit similar performance to NS in average latency.

\textbf{Latency Distribution.}
%As follows, we dived into the detail latency performance in time (tick) series.
\figref{fig:256} plots the latency traces on a 1000MB dataset with $uf$=256K per tick.
Only the latency traces between the 150th and 350th ticks are plotted, which include a full checkpointing period.
Other fragments of the traces show a similar pattern.
\figref{fig:256-part} zooms in on the details of the differences between Fork, ZZ, HG, and PB.

We observe that the latency of each algorithm is relatively stable.
A latency spike usually appears at the beginning of a checkpointing when the snapshotter enters the snapshot taken phase.
NS, COU, Fork and ZZ show notable latency spikes because of the O(n) time complexity.
%\sss{It indicates that they are more volatile than the others, especially for NS. }
The dramatic latency spikes of NS can cause trouble in practical applications.

%\tabref{tlb:var} shows the latency variances with $uf$ 16K and 128K per tick, respectively.
%\TODO{one sentence describing the results in the table}
%\TODO{one summary sentence comparing the performance of latency spikes and variance of the algorithms}
%\xxx{DELETE variance part.}
%\begin{table}[!htb]
%	\centering
%	\caption{Latency variance under different \textit{uf}}
%	\tabcolsep0.02in
%	\label{tlb:var}
%	\newcolumntype{L}[1]{>{\raggedright\arraybackslash}p{#1}}
%	\newcolumntype{C}[1]{>{\centering\arraybackslash}p{#1}}
%	\newcolumntype{R}[1]{>{\raggedleft\arraybackslash}p{#1}}
%	\begin{tabular}{|c|c|c|c|c|c|c|c|}
%		\hline
%		$uf$ & NS & COU & Fork & ZZ & PP & \ll & \mk   \\
%		\hline
%		16k  &  $2.33\times 10^8$	& 5634.0 &  7166.1  &	5520.0	& 370.1	& 109.4 & 111.3	  \\
%		\hline
%		128k  & $2.27\times 10^8$	& 7929.4 &	20870  & 28033.7	& 9269.1 &	3387.3	& 4699.8   \\
%		\hline
%	\end{tabular}
%\end{table}

\textbf{Maximum Latency.}
The taken phase time of the snapshotter dominates the maximum latency of the client thread.
\figref{fig:size:prepare} shows the maximum latency with the increase in dataset size.
The update frequency is set to 256K per tick in this experiment.
We can observe that the maximum latency of PP, \ll, and \mk are several orders of magnitude lower than that of NS, COU, Fork, and ZZ.
Moreover, the maximum latency of the latter algorithms becomes steadily larger with the increase in dataset size.
The good performance of PP, HG and PB is due to the pointer swapping technique.

\figref{fig:uf:prepare} further shows the impact of $uf$ on the maximum latency on a 1000MB dataset.
All curves remain horizontal because the maximum latency is only influenced by data size.

In sum, comparing average latency, latency distribution and maximum latency, our improved algorithms, \ll and \mk, exhibit better performance than other algorithms.

\begin{figure*}[!htb]
	\centering
	\begin{minipage}[t]{0.32\textwidth}
		\centering
		\includegraphics[width=\textwidth]{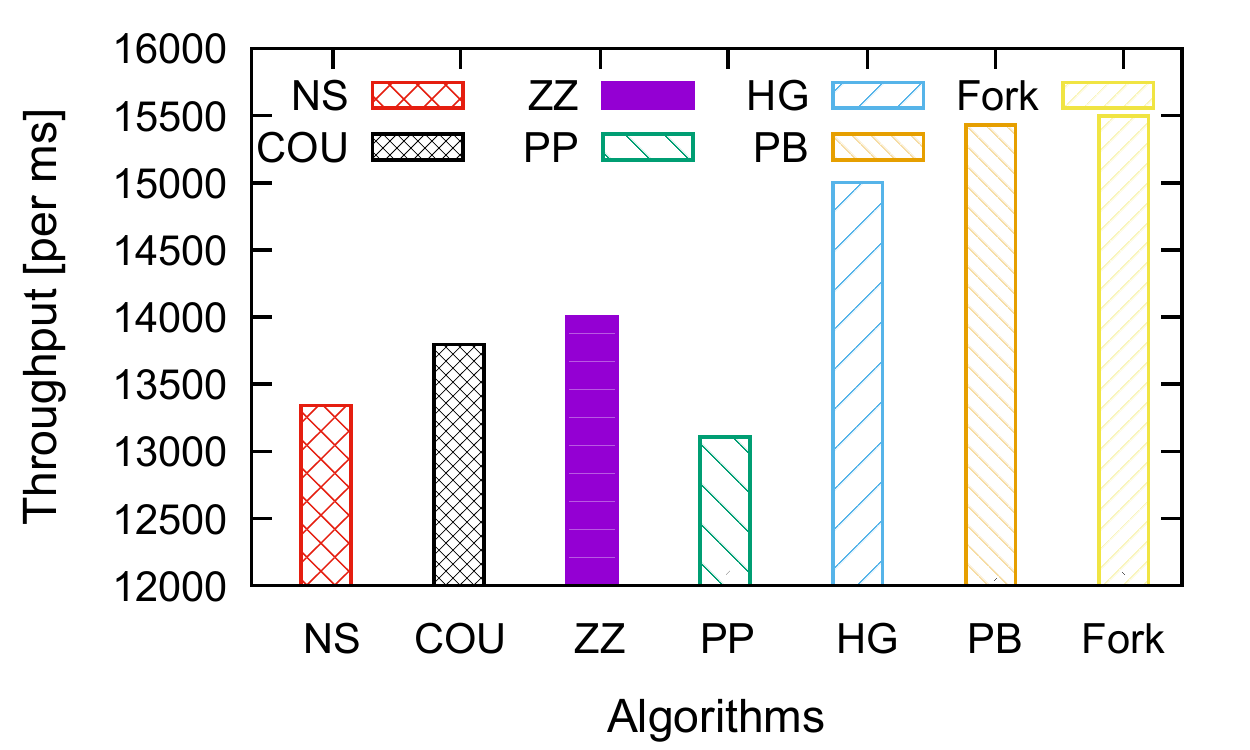}
		\caption{Maximum throughput}
		\label{fig:tps}
	\end{minipage}
	\hspace{0.1cm}
	\begin{minipage}[t]{0.32\textwidth}
		\centering
		\includegraphics[width=\textwidth]{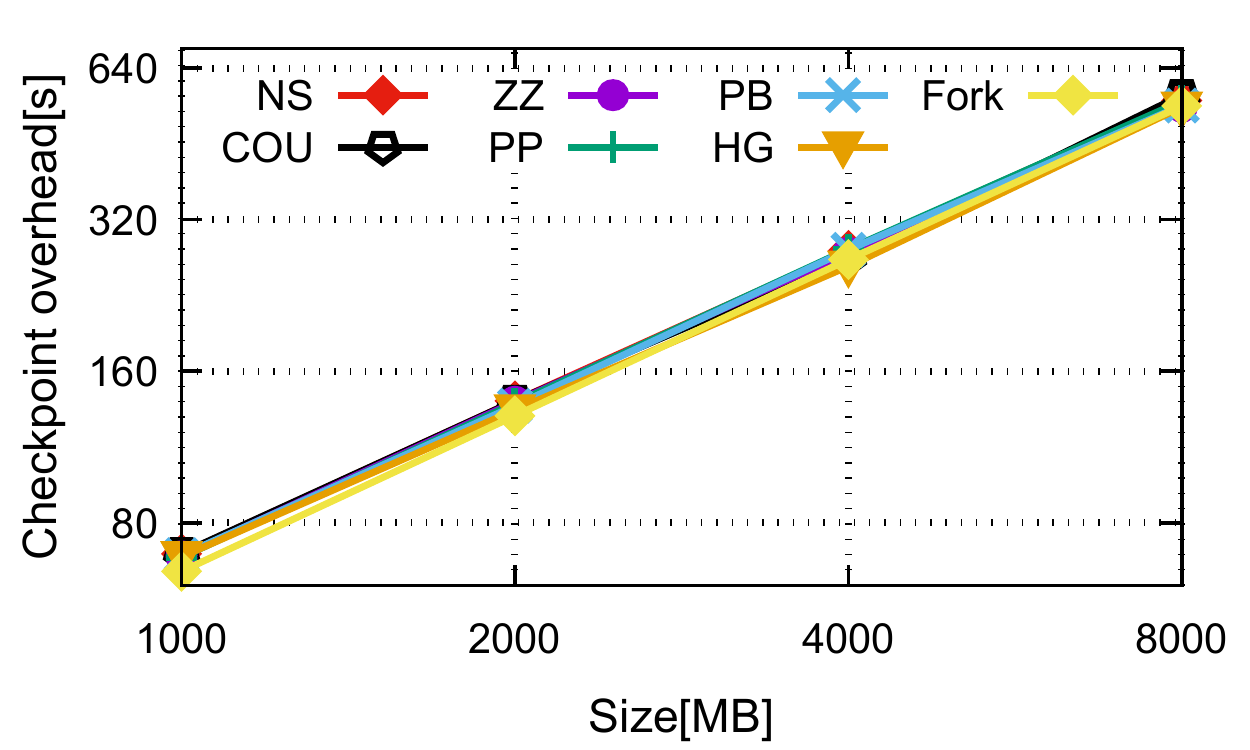}
		\caption{Data size vs. Checkpointing overhead}
		\label{fig:size:overhead}
	\end{minipage}
	\hspace{0.1cm}
	\begin{minipage}[t]{0.32\textwidth}
		\centering
		\includegraphics[width=\textwidth]{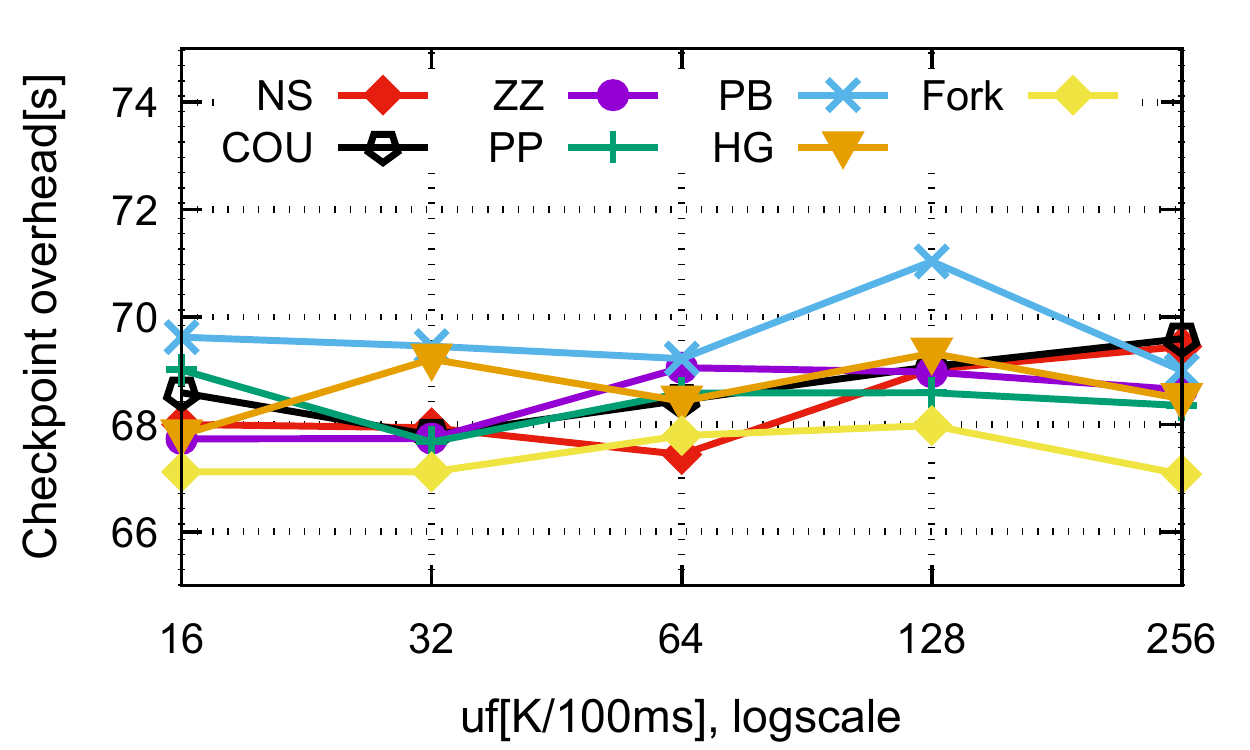}
		\caption{\textbf{\textit{uf}} vs. Checkpointing overhead}
		\label{fig:uf:overhead}
	\end{minipage}
\end{figure*}

\textbf{Maximum Throughput.}\label{SubSec:synMaxThroughput}
Maximum throughput is one metric to assess the maximum load capacity.
\figref{fig:tps} shows the maximum throughput per millisecond on a 8000MB dataset.
Unlike the previous experiments, the length of the idle stage in a tick is set to zero.
Thus, the client can update as fast as possible in a full-update (no-wait) mode.
Further, we turn off the Dump operation to observe the throughput limit.

\ll, \mk and Fork exhibit better performance in maximum throughput.
In fact, the maximum throughput is a comprehensive reflection of the results in \figref{fig:uf:latency} and \figref{fig:uf:prepare}.
For instance, although NS has the shortest average latency, its latency spike is the most obvious, which leads to a bad maximum throughput.

\textbf{Checkpointing Overhead.}\label{SubSec:synOverhead}
Checkpoint overhead is the traverse and dump overhead of the the snapshotter thread.
All algorithms perform a full snapshot for fair comparison.
For incremental snapshot algorithms, a full snapshot is obtained by merging the incremental dumped data with the last snapshot.
This can be achieved by using the \textit{Copy} and \textit{Merge} proposed by \cite{Cao.13}.
\textit{Merge} is more efficient than \textit{Copy} in terms of the memory maintenance cost.
Therefore, for Algorithm \ref{alg:LL} Line 7 of Snapshotter::TraverseSnapshot(), we apply \textit{Merge} to construct a new full snapshot.

\figref{fig:size:overhead} shows the trend in overhead on varying dataset sizes.
The update frequency is fixed to 256K per tick.
The checkpointing overheads of all the algorithms increase linearly, and there is little difference between their overheads;
the overheads are primarily dominated by the dataset size written to the external memory.
\figref{fig:uf:overhead} shows the overheads with different update frequencies on a dataset of 1000MB.
The overheads remain almost constant across all the algorithms because the dataset size is fixed.

\subsubsection{Does Fork really good enough?}
\begin{figure}[h]
	\centering
	\includegraphics[width=0.32\textwidth]{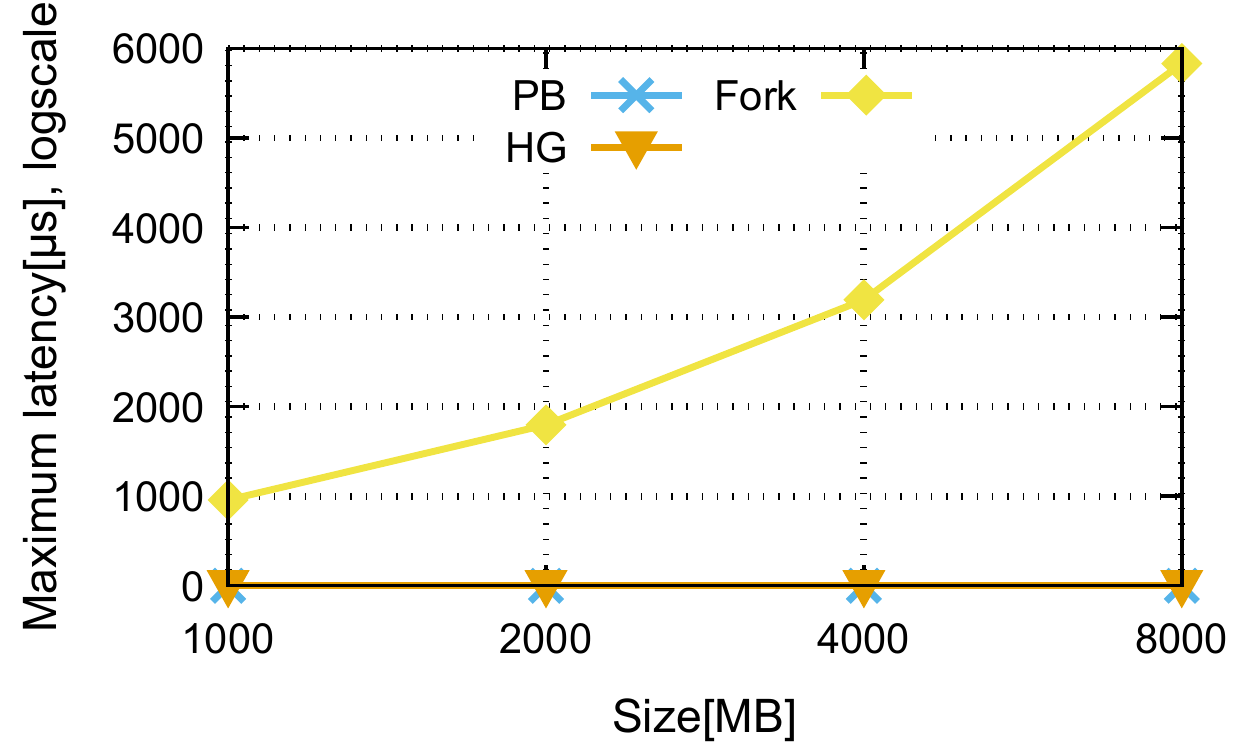}\\
	\caption{fork performance}
	\label{fig:fork}
\end{figure}

\figref{fig:fork} shows the performance comparison of Fork, HG, and PB with data size ranging from 1000~MB to 8000~MB. In particular, the results reflect the limitations of Fork.
It is no longer comparable with \ll and \mk in the case of maximum latency.
As stated in~\cite{redis-latency}, although it is an OS kernel function, fork is still expensive on most Unix-like systems.
The costs are the result of copying the memory page table from the parent process to the child process.
Let the memory page size be 4 KB for a Linux x86\_64 system whose pointer size is 8 Bytes.
Thus, to address a 50 GB memory space, the size of the memory page table should be $\frac{50 GB}{4 KB}$$\times$8B$\approx$ 100 MB.
This will result in the allocation and copying of 100 MB memory whenever a checkpoint occurs.
Suppose that the memory bandwidth is 2.5 GB/s; the latency spike is about 40 ms, which accords with the results in \figref{fig:size:prepare} and \figref{fig:fork}.

%\figref{fig:fork} gives th the performance of fork, HG and PB  between datasize from 1GB to 8GB.

\textbf{Summary.}
Our benchmark evaluations on a synthetic workload reveal the following findings.
\begin{itemize}	
  \item
  For applications where only backend performance is a concern, the snapshot algorithms should have low average latency.
  NS, Fork, \ll and \mk all are applicable (see \figref{fig:uf:latency} and \figref{fig:size:latency}).
  \item
  For interaction-intensive applications (\ie frequent updates), latency spikes should be included to assess the snapshot algorithms.
  PP, HG and PB outperform the others in terms of the number of spikes and the value of spikes (\ie maximum latency), while NS performs the worst (see \figref{fig:256}, \figref{fig:size:prepare} and \figref{fig:uf:prepare}).
  \item
  Fork outperforms NS, COU, ZZ and PP in terms of both latency and throughput (see \figref{fig:256}, \figref{fig:size:prepare} and \figref{fig:tps}); in addition, fork has a simple engineering implementation; fork is therefore adopted in several industrial IMDBs such as Redis.
  \item
  The latency performances of PB and HG are not affected by the data size (see \figref{fig:size:prepare} and \figref{fig:fork}).
  In general, PB and HG are more scalable than the other algorithms including fork.
  \item
  NS, Fork, COU, ZZ and PP are fit for specific applications (\ie they perform well either on latency or throughout).
  PB and HG trade off latency, throughput and scalability, which are fit for a wider range of applications.	
\end{itemize}

\subsection{Benchmark study of Virtual Snapshot}\label{SubSec:extend_exp}
% 这部分，我们主要测试的是在不同并发线程数量下，CALC，PB，HG算法分别对应的每ms内的平均吞吐率。
% 图。。展示了在100MB数据量下，线程数从1到1024的吞吐率对比图。
The above experiments were conducted under the update intensive physical snapshot scenario. In this section, we present a comparison of CALC, vHG and vPB under the virtual snapshot scenario. The transaction concurrent control method used here is the general Strict two-phase Locking Protocol (S2PL)~\cite{bernstein1987rrency} as in~\cite{ren2016low-overhead}. Note that any other concurrent control methods (e.g., MVCC, OCC, etc.) are also applicable as long as they are under the same concurrency control protocol.

\begin{figure*}[!htb]
	\centering
	\begin{minipage}[t]{0.32\textwidth}
		\centering
		\includegraphics[width=\textwidth]{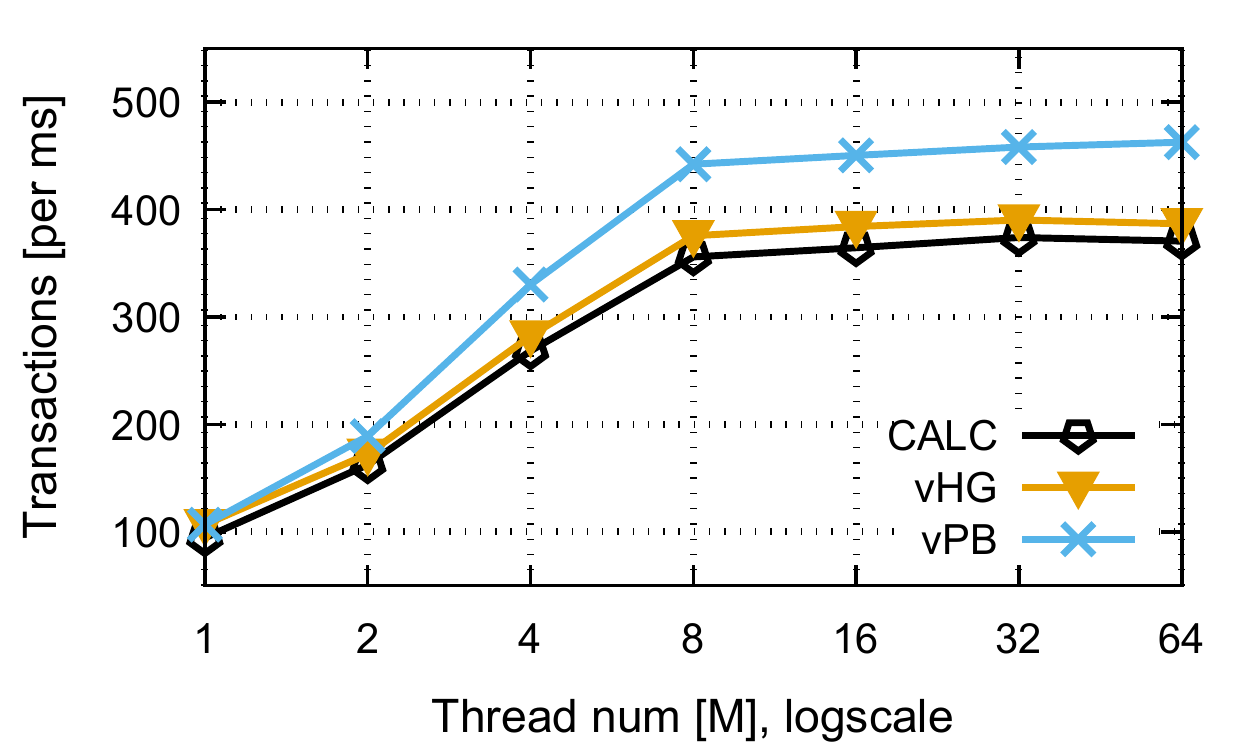}
		\caption{Thread num \textbf{vs.} Transaction throughput}
		\label{fig:conccurent}
	\end{minipage}
	\hspace{0.1cm}
    \begin{minipage}[t]{0.32\textwidth}
		\centering
		\includegraphics[width=\textwidth]{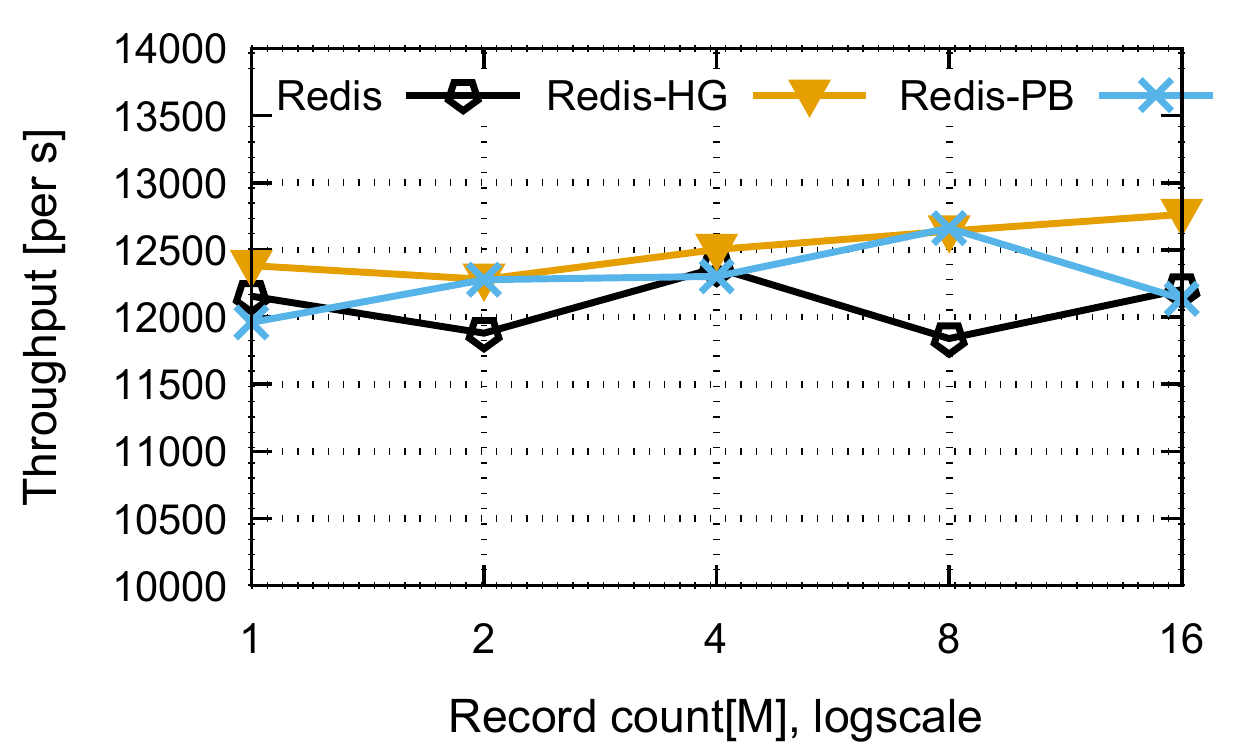}
		\caption{Redis: YCSB record count vs. Throughput }
		\label{fig:redis_tps}
	\end{minipage}
	\hspace{0.1cm}
	\begin{minipage}[t]{0.32\textwidth}
		\centering
		\includegraphics[width=\textwidth]{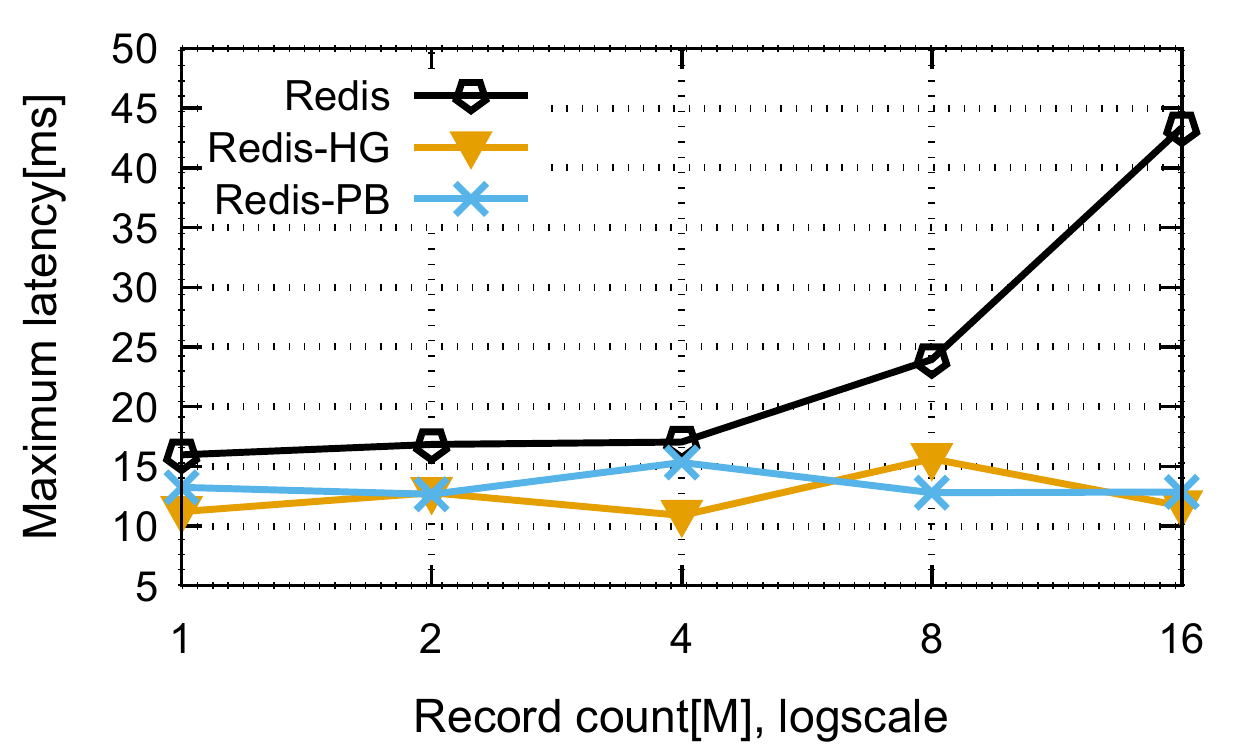}
		\caption{Redis: YCSB record count \textbf{vs.} Maximum latency}
		\label{fig:redis_spike}
	\end{minipage}    
\end{figure*}

\figref{fig:conccurent} compares the throughput of CALC, vHG, and vPB on a 100MB dataset.
Since vHG and vPB  do not require page copy operations, they have greater workload capacity than CALC for all the multi-thread cases.
Interestingly, we observe that the throughput tends to be stable when the thread number is larger than 8.
%With more number of threads in the system, the performance is not always better.
Even if the number of threads in the system continues to increase, the performance will not always improve.
This phenomenon can be explained by the heavy lock contentions among threads.
The result is similar to that of the DBx1000 project~\cite{DBLP:journals/pvldb/YuBPDS14}.

\subsection{Performance in Industrial IMDB System}\label{Sec:SystemValidations}
%\subsection{Performance Study on End-to-End Prototype}
Redis is a popular In-Memory NoSQL system and it utilizes fork() to persist data~\cite{redis-persistency}.
To generate the persistent image (a.k.a. RDB file) in the background, Redis has to invoke the system call \textit{fork}() to start a child process to execute snapshot and dumping work.
From the above benchmark study, we see that fork() indeed performs better than mainstream snapshot algorithms including NS, COU, ZZ and PP in terms of latency and throughput.
However, we also suspect that fork() will incur dramatic latency on large datasets, which limits the scalability of Redis.
In fact, database users usually restrain the data size of a running Redis instance in practice~\cite{ec2-redis}.
In this performance study, we aim to harness proper snapshot algorithms to improve the scalability of snapshots in Redis.

\begin{figure*}[!htb]
	\centering	
	\begin{minipage}[t]{0.32\textwidth}
		\centering
		\includegraphics[width=\textwidth]{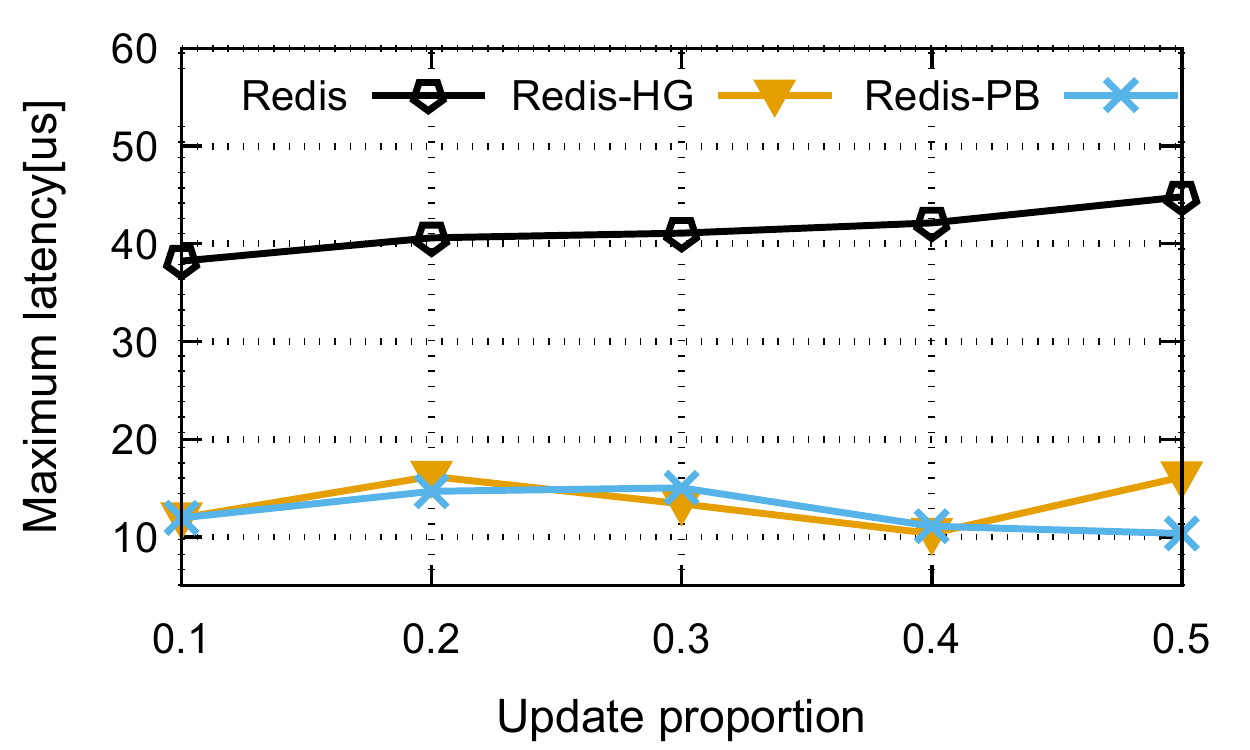}
		\caption{Redis: Update proportion \textbf{vs.} Maximum latency}
		\label{fig:redis_rw}
	\end{minipage}
	\hspace{0.1cm}
	\begin{minipage}[t]{0.32\textwidth}
		\centering
		\includegraphics[width=\textwidth]{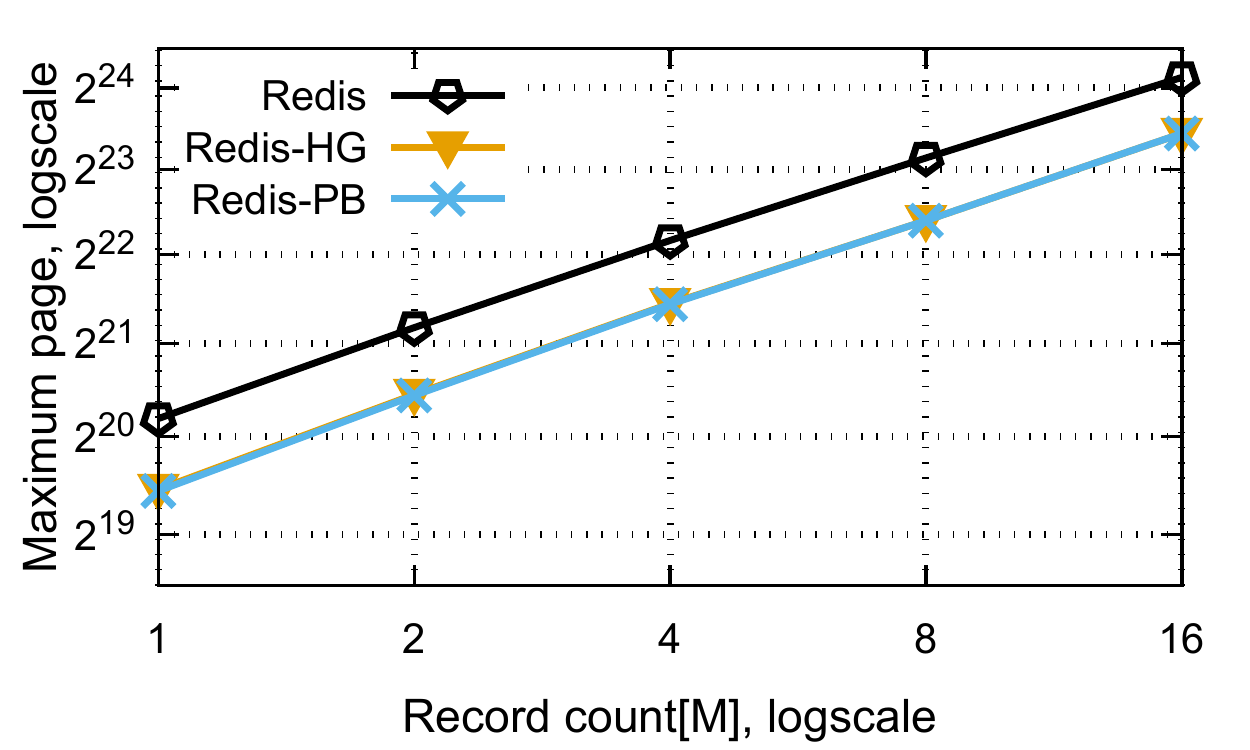}
		\caption{Redis: YCSB record count \textbf{vs.} Maximum memory cost}
		\label{fig:redis_mem}
	\end{minipage}
	\hspace{0.1cm}
	\begin{minipage}[t]{0.32\textwidth}
		\centering
		\includegraphics[width=\textwidth]{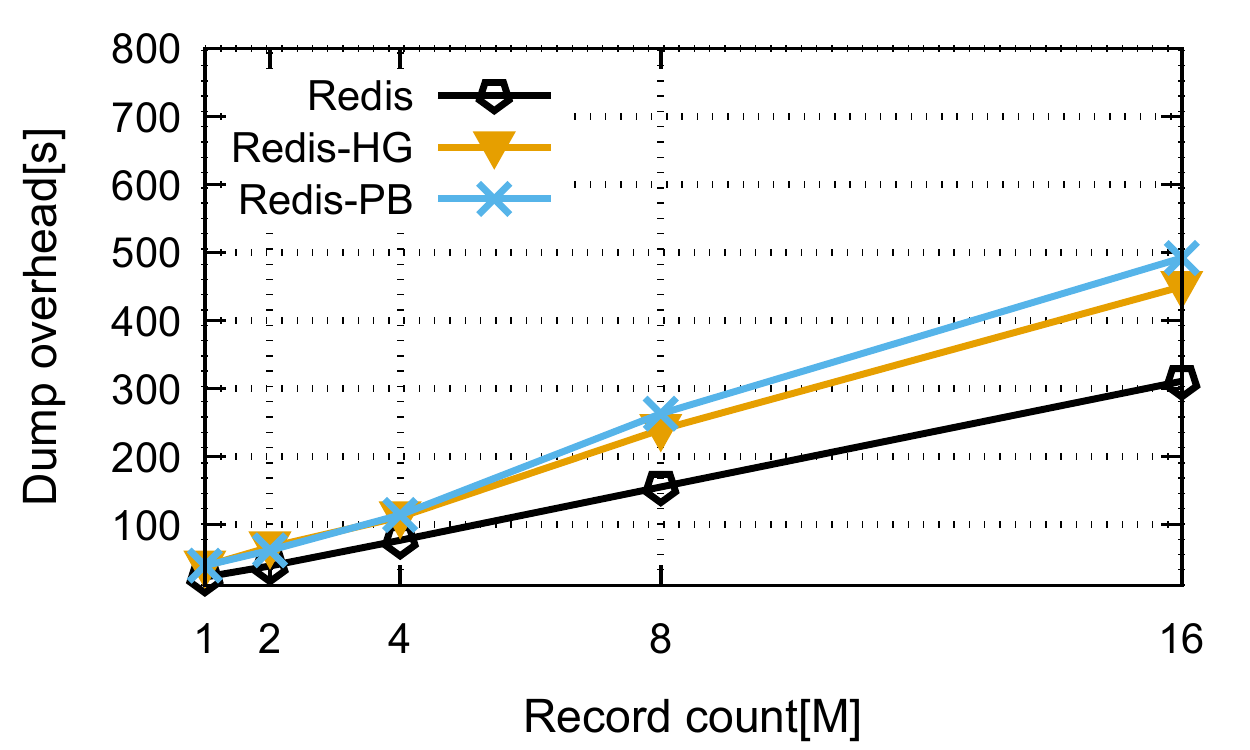}
		\caption{Redis: YCSB record count \textbf{vs.} Dump overhead}
		\label{fig:redis_dump}
	\end{minipage}
\end{figure*}

%Two prototypes, \textbf{Redis-HG} and \textbf{Redis-PB}, are implemented based on Redis.
%We first briefly describe the implementations of the Redis version algorithms in Section \ref{subsec:redis} .
%And then with the YCSB benchmark workloads, the evaluation results on throughput, maximum latency, maximum memory footprint and dump overhead are presented.
%The results show that our algorithms have significant performance improvement compared to the official Redis.

\subsubsection{Snapshot Algorithm Selection}\label{subsec:redis}

Fork has weak scalability due to its O(n) time complexity.
It is posted in the official website~\cite{redis-persistency} that ``\textit{Fork() can be time consuming if the dataset is large, and as a result, Redis may stop serving clients for milliseconds or even one second if the dataset is large and the CPU performance not great}''.

To optimize the scalability of Redis, an option is to replace fork with snapshot algorithms of O(1) complexity.
Here, we implement two Redis variants Redis-HG and Redis-PB using the HG and PB algorithms, respectively.
Both variants are a single-process with double-thread (client and snapshotter).
We do not choose ZZ, for it is only suitable for small datasets;
during the taken phase, ZZ needs to operate all the bit flags.
Traversing all the keys is time consuming and is almost equal to executing the ``keys *" directive.
%What's more, its feasibility is denied by Hyper \cite{muhe2011efficiently}, a high-performance in-memory database. \TODO{what do you mean by ``its feasibility is denied''} \xxx{DELETE}
%Actually, we have implemented it, but its performance are very terrible, for the space reason, we discard Redis-ZZ prototype.
PP is also excluded due to the three copies of the memory footprint.
In addition, the Redis architecture is unfit for integrating the PP algorithm.

Note that, as a key-value database, Redis internally maintains a large hash table in the main memory.
Alternatively, for each key, we build the array data model as its corresponding value.
Although the arrays in the model are physically broken up into fragments, they are still logically traversable through the hash table.
To save memory (see \secref{sec:comp}), we introduce the \textit{garbage collection} and \textit{dynamic memory allocation} techniques.

\subsubsection{Dataset}
%Since TPC-C is not suitable for benchmarking NoSQL databases, we chose YCSB (Yahoo! Cloud Serving Benchmark) \cite{Cooper.14} to validate the practical performance of our algorithms in real big data application scenarios.
%We exhibit the experimental results on YCSB to show the improvements of our variants to the official Redis during  persistence period.
We chose YCSB (Yahoo! Cloud Serving Benchmark)~\cite{Cooper.14} to validate the practical performance of our prototype.
A workload of YCSB is a dataset plus a set of read and write operations, \ie, a transaction set.
The dataset is loaded into the database and then consumed by the transaction set.
YCSB predefines six main workloads.
We implement our own workloads based on \textit{Workload A} (update heavy workload) and \textit{Workload B} (read mostly workload), which accord with our aim to evaluate performance for frequent consistent checkpointing.
\tabref{tlb:setup} shows the detailed workload setups.
Those defined by YCSB are in bold.

The Redis configuration file ``\textit{redis.conf}'' contains a number of directives.
We use directive ``\textit{save 10 1}" to configure Redis to automatically dump the dataset to disk every 10 seconds if there is at least one change in the dataset.

\begin{table}[htb]
	\centering
	\caption{Parameters of YCSB workloads}
	\label{tlb:setup}
	\begin{tabular}{|c|c|}
		\hline
		{\bfseries Parameters} &   {\bfseries Setting} \\
		\hline
		% Transition thread &          1 \\
		Loading thread &          256 \\
		\hline
		Distribution &       Zipfian \\
		\hline
		Operation count &    4M \\
		\hline
		Update proportion &        \textbf{0.1}, 0.2, 0.3, 0.4, 0.5 \\
		\hline
		Record count & \textbf{1M}, 2M, 4M, 8M, 16M \\
		\hline
	\end{tabular}
\end{table}

\subsubsection{Performance}
We mainly evaluate the performance of the two Redis variants in terms of throughput, latency, effect of updates, memory and snapshot overhead.

\textbf{Throughput.}
\figref{fig:redis_tps} illustrates the change of throughput as the benchmark record count grows.
The trends are consistent with those shown in \figref{fig:size:latency}.
We make two observations.
\textit{(i)} The throughput of all the algorithms is insensitive to the dataset size.
\textit{(ii)} Redis-HG and Redis-PB have similar throughput performance to the default Redis;
although Redis-HG and Redis-PB can avoid locks between data updating and dumping, they need additional checking through the hash table for each read/write operation.
Therefore, the throughout improvement is marginal.

\textbf{Maximum Latency.}
\figref{fig:256} compares the differences in latency spikes and \figref{fig:size:prepare} shows the maximum latency with the increase of the dataset size.
\figref{fig:redis_spike} plots the maximum latency with the record count from 1 million to 16 million, approximately up to 50GB (with update proportion = 0.1).

The default Redis incurs a dramatic increase in maximum latency when the record count reaches 8 million.
This result is consistent with the Redis document for which the maximum latency becomes huge because of the invocation of fork().
Redis-\mk and Redis-\ll have similar maximum latencies, and both remain stable with the growth of the record count.
This can be explained by the pointer swapping technique employed in the snapshot taken phase, which only needs to be almost constant and incur a small cost.
We expect that the maximum latency of official Redis implementation will grow rapidly with the record count until eventually quiescing the system, which leads to weak scalability.
Conversely, Redis-HG and Redis-PB can scale to larger datasets than the default Redis.

\textbf{Effects of Updates on Latency.}
\figref{fig:redis_rw} shows the maximum latency with fixed record counts of 8 million and a varying proportion of updates from 0.1 to 0.5.
As shown, the maximum latency of Redis grows with more updates, while those of Redis-HG and Redis-PB still remain relatively stable.
We conclude that Redis-HG and Redis-PB are more suitable for update-intensive applications.

\textbf{Maximum Memory Footprint.}\label{SubSec:MemoryFootPrint}
Since the default Redis persistence strategy depends on forking a child process to dump the snapshot, the additional application dataset size of the memory footprint is inevitable.
Although at the beginning of a fork, the parent and child processes share a single data region in memory, the actual size of the memory consumed will increase with frequent data updates (\ie page duplication).
That is, the memory footprint depends on the workload.
In the worst case (update intensive), fork will lead to a memory spike (almost double the memory footprint)~\cite{redis-mem-spike}.
As explained in Redis FAQ~\cite{redis-faq}, the fork may fail when the Redis memory size is larger than half of system memory.
Although the fork failure can be avoided by setting parameter $overcommit\_memory$ to 1, there still exists the risk of being killed by the OS' OOM killer.
Based on experience, the case where the redis instance is larger than half of the local physical memory is dangerous.

In principle, Redis-HG and Redis-PB need a memory footprint that is twice the size.
To reduce memory usage, we leverage the \textit{dynamic memory allocation} and the \textit{garbage collection}  technologies.
Once a value has been dumped to the disk and the value is not up-to-date, the corresponding value portion should be identified as garbage that can be destroyed and reused by the system now or later.
\figref{fig:redis_mem} shows the comparison of the maximum memory cost.
All comparisons are linear to the dataset size.
The memory cost of Redis-HG and Redis-PB are similar and far smaller than the original Redis.

\textbf{Checkpointing Overhead.}
\figref{fig:redis_dump} presents the checkpointing (\ie RDB) overhead.
The results are similar to those in \figref{fig:size:overhead}.
The scale of the record count ranges from 1 million to 16 million.
The checkpointing overhead grows linearly with the dataset size.
For small datasets, the dump overheads of Redis-HG and Redis-PB are close to that of the original Redis.
The gap increases slowly with the increase in dataset size.
Note that the two variants need additional state checking to determine the appropriate copy of data for dumping while the default Redis' child thread only needs to traverse the hash table to flush all the key-value pairs.
Fortunately, the double-thread design effectively separates the updating and dumping tasks and induces only a slightly longer background dumping period.
Furthermore, the overhead gap can be reduced by leveraging high-speed disks and large memory buffers.

\textbf{Summary.}\label{sec:summary}
Redis with the built-in fork() function is unscalable (see \figref{fig:redis_spike}).
By replacing the default fork with HG and PB, the two variants, Redis-HG and Redis-PB, exhibit better scalability.

\section{Conclusions}\label{Sec:Conclusions}

In this paper, we analyze, compare, and evaluate representative in-memory consistent snapshot algorithms from both academia and industry.
Through comprehensive benchmark experiments, we observe that the simple fork() function often outperforms the state-of-the-arts in terms of latency and throughput.
However, no in-memory snapshot algorithm achieves low latency, high throughput, small time complexity, and no latency spikes at the same time; however, these requirements are essential for update-intensive in-memory applications.
We propose two lightweight improvements over existing snapshot algorithms, which demonstrate better tradeoff among latency, throughput, complexity and scalability.
We implement our improvements on Redis, a popular in-memory database system.
Extensive evaluations show that the improved algorithms are more scalable than the built-in fork() function.
We have made the implementations of all algorithms and evaluations publicly available to facilitate reproducible comparisons and further investigation of snapshot algorithms.

%In Figure \ref{fig:model}, we broadly classify typical checkpointing algorithms based on two dimensions, i.e., the \textit{consistency} and the \textit{concurrency}.
%\begin{figure}[htb]
%	\centering
%	\includegraphics[width=0.5\textwidth]{fig/class.pdf}\\
%	\caption{Consistent checkpointing running model}
%	\label{fig:model}
%\end{figure}

% or


\section{Future work}
This work discusses leveraging snapshot algorithms to perform checkpoints.
As described in \secref{sec:introduction}, consistent snapshots are not only
used for consistent checkpoints but are also employed in HTAP systems~\cite{OzcanTT17,stonebraker2007end,kemper2011hyper,plattner2009common,lang2016data,funke2011benchmarking,Meng2017,farber2012sap,sikka2013sap},
\ie Hyper, HANA and SwingDB.
Although there are many studies about concurrency control protocols for OLTP systems~\cite{DBLP:journals/pvldb/YuBPDS14,ren2012lightweight,DBLP:conf/sosp/TuZKLM13,DBLP:conf/sigmod/YuPSD16,DBLP:journals/pvldb/WuALXP17,DBLP:conf/sigmod/LimKA17}, few works address HTAP's concurrency control; thus, we plan to build a prototype based on snapshot concurrency control to fill this gap in the future.

% or


%\appendices

%\section{}

\ifCLASSOPTIONcompsoc
  \section*{Acknowledgments}
\else
  \section*{Acknowledgment}
\fi

The authors would like to thank Wenbo Lang, Phillip Saenz and the anonymous reviewers.
Guoren Wang is the corresponding author of this paper.
Lei Chen is supported by the Hong Kong RGC GRF Project  16214716 , National Grand Fundamental Research 973 Program of China under Grant 2014CB340303, the National Science Foundation of China (NSFC) under Grant No. 61729201, Science and Technology Planning Project of Guangdong Province, China, No. 2015B010110006, Webank Collaboration Research Project, and Microsoft Research Asia Collaborative Research Grant.
Guoren Wang is supported by the NSFC (Grant No. U1401256, 61732003, 61332006 and 61729201).
Gang Wu is supported by the NSFC (Grant No. 61370154).
Ye Yuan is supported by the NSFC (Grant No. 61572119 and 61622202) and the Fundamental Research Funds for the Central Universities (Grant No. N150402005).

\ifCLASSOPTIONcaptionsoff
  \newpage
\fi



\bibliographystyle{IEEEtran}
\bibliography{IEEEabrv,list}


\begin{IEEEbiography}[{\includegraphics[width=1in,height=1.25in,clip,keepaspectratio]{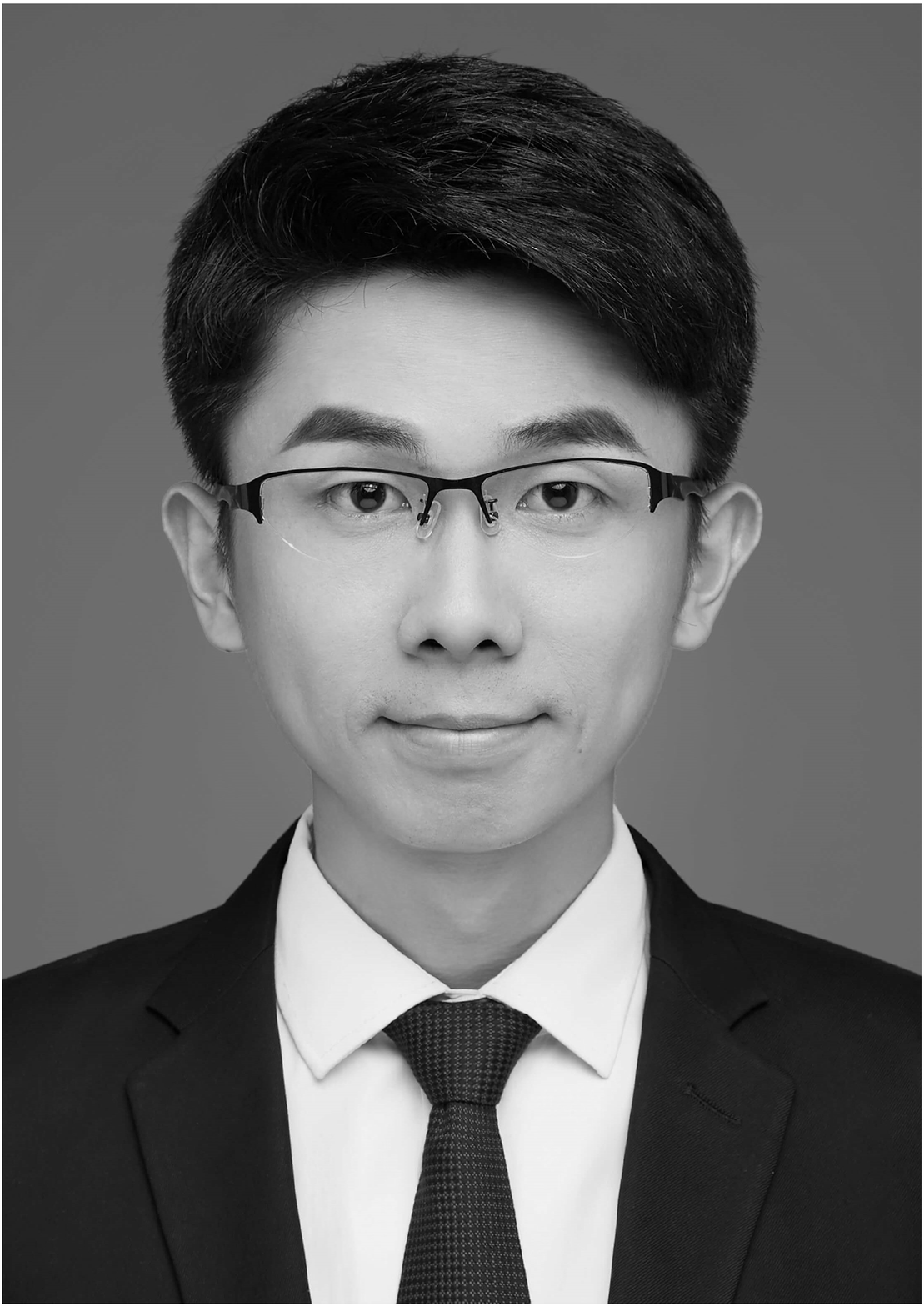}}]{Liang Li}
received his BSc degree from the College of Computer Science and Engineering of Northeastern University, China in 2014. Currently, he is a PhD student in Computer Science and Engineering at Northeastern University. His main research interests include in-memory database systems, distributed systems, and database performance.
\end{IEEEbiography}
\vspace{-1cm}
\begin{IEEEbiography}[{\includegraphics[width=1in,height=1.25in,clip,keepaspectratio]{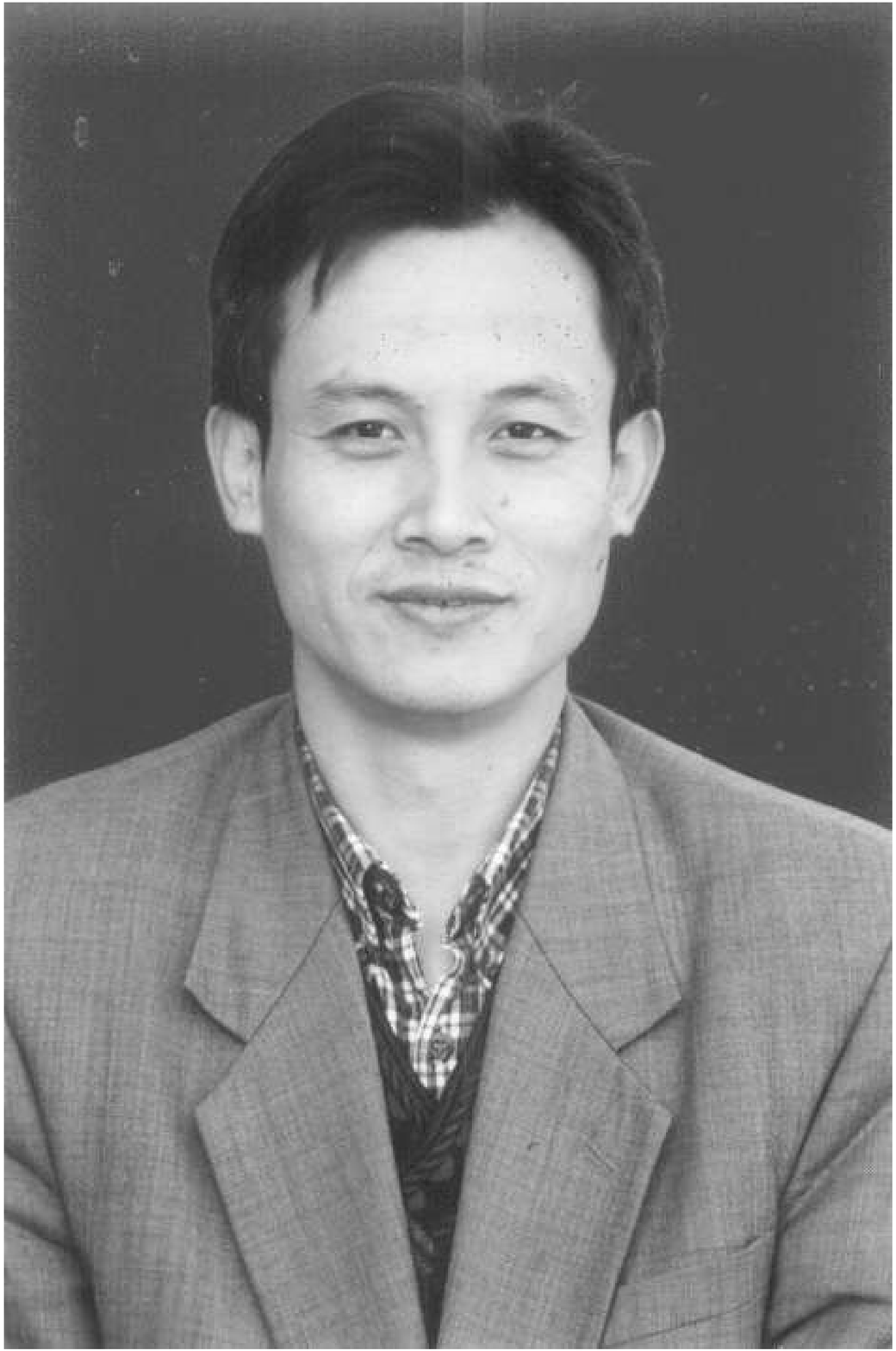}}]{Guoren Wang}
received his BSc, MSc and PhD degrees in Computer Science from Northeastern University, China in 1988, 1991 and 1996, respectively. Currently, he is a Professor in the Department of Computer Science at Beijing Institute of Technology, China. His research interests include XML data management, query processing and optimization, bioinformatics, high-dimensional indexing, parallel database systems, and P2P data management.
\end{IEEEbiography}
\vspace{-1cm}
\begin{IEEEbiography}[{\includegraphics[width=1in,height=1.25in,clip,keepaspectratio]{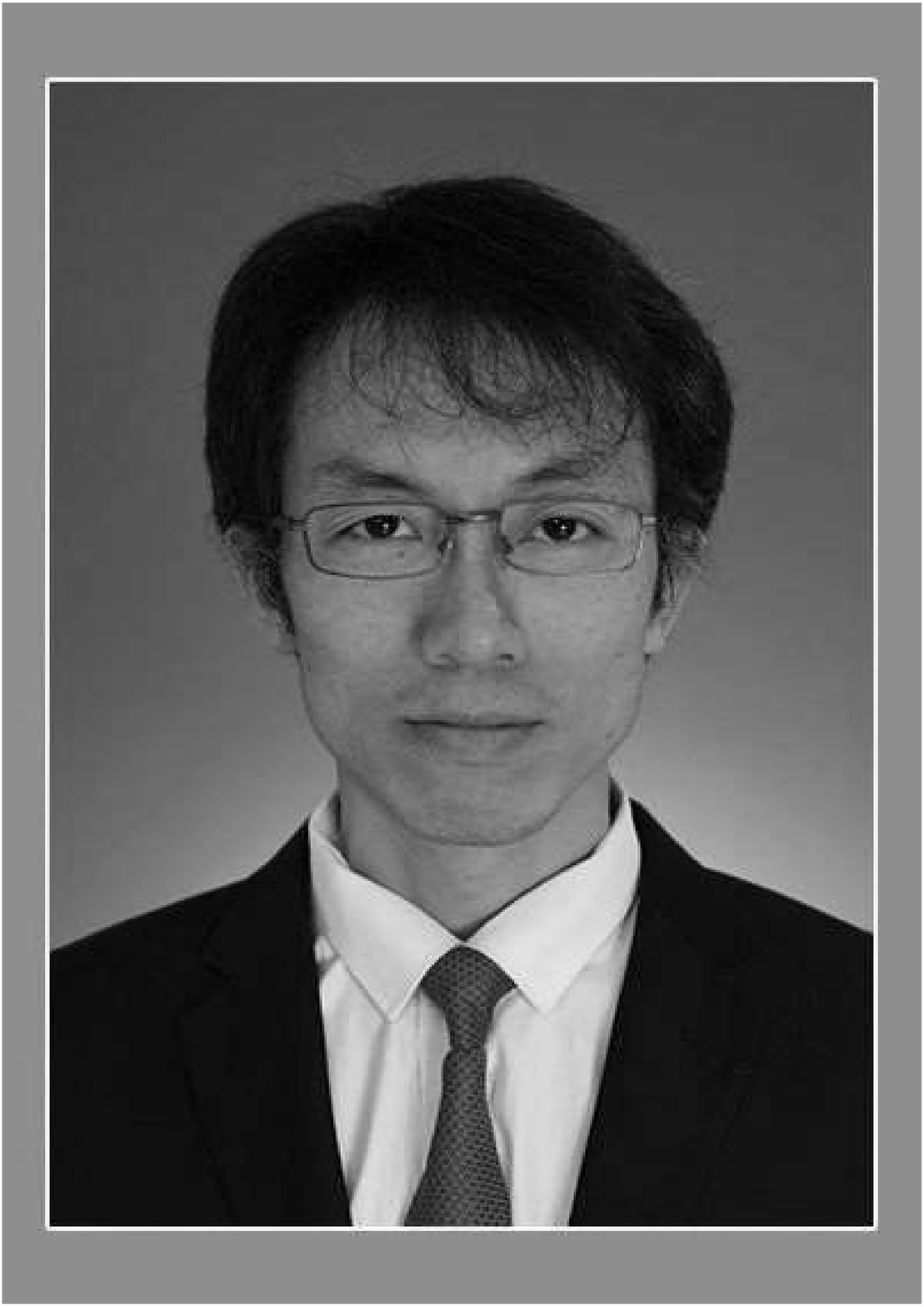}}]{Gang Wu}
received his BS and MS degrees in Computer Science from Northeastern University, China in 2000 and 2003, respectively, and his PhD degree in Computer Science from Tsinghua University in 2008.
He is now an Associate Professor at the College of Information Science and Engineering at Northeastern University.
His research interests include in-memory databases, graph databases, and knowledge graphs.
\end{IEEEbiography}
\vspace{-1cm}
\begin{IEEEbiography}[{\includegraphics[width=1in,height=1.25in,clip,keepaspectratio]{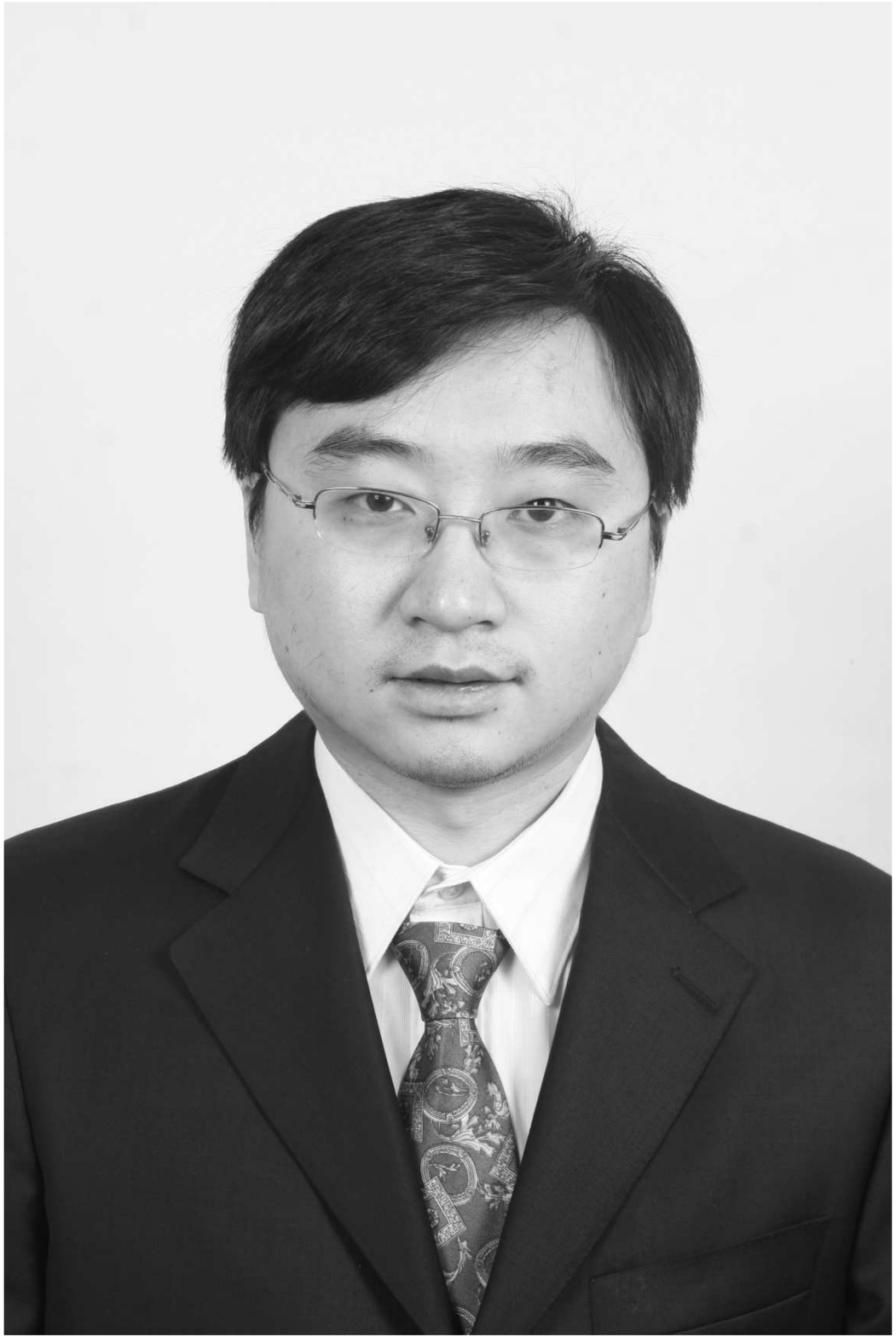}}]{Ye Yuan}
received his BS, MS and PhD degrees in Computer Science from Northeastern University, China in 2004, 2007 and 2011,
respectively. He is now a Professor at the College of Information Science and Engineering at Northeastern University. His research interests include graph databases, probabilistic databases, data privacy-preserving and cloud computing.
\end{IEEEbiography}
\vspace{-1cm}
\begin{IEEEbiography}[{\includegraphics[width=1in,height=1.25in,clip,keepaspectratio]{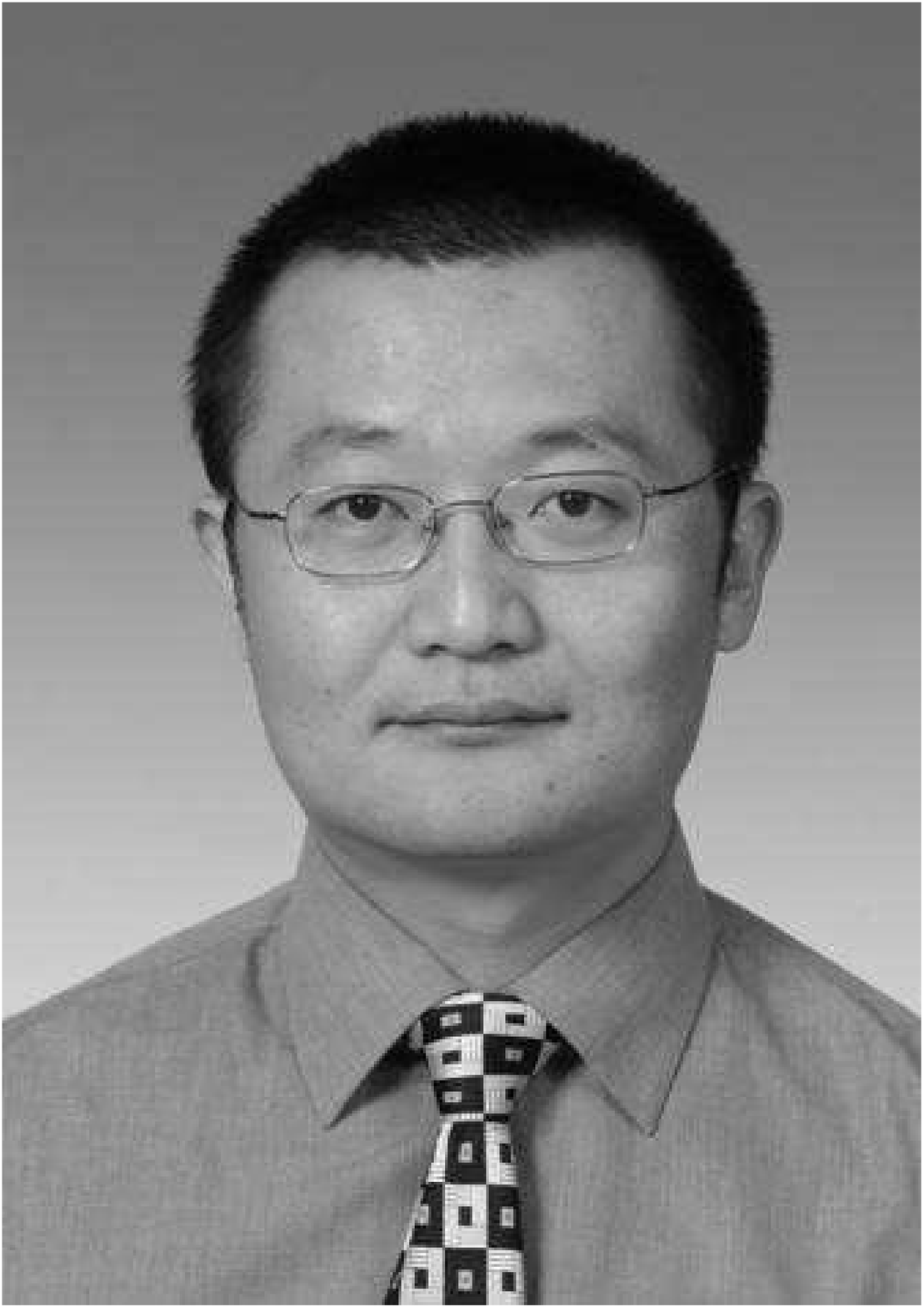}}]{Lei Chen}
received his BS degree in Computer Science and Engineering from Tianjin University, China in 1994, his MA degree from the Asian Institute of Technology, Bangkok, Thailand in 1997, and his PhD degree in Computer Science from the University of Waterloo, Canada in 2005. He is currently an Associate Professor in the Department of Computer Science and Engineering, Hong Kong University of Science and Technology. His research interests include crowdsourcing over social media, social media analysis, probabilistic and uncertain databases, and privacy-preserved data publishing.
\end{IEEEbiography}
\vspace{-1cm}
\begin{IEEEbiography}[{\includegraphics[width=1in,height=1.25in,clip,keepaspectratio]{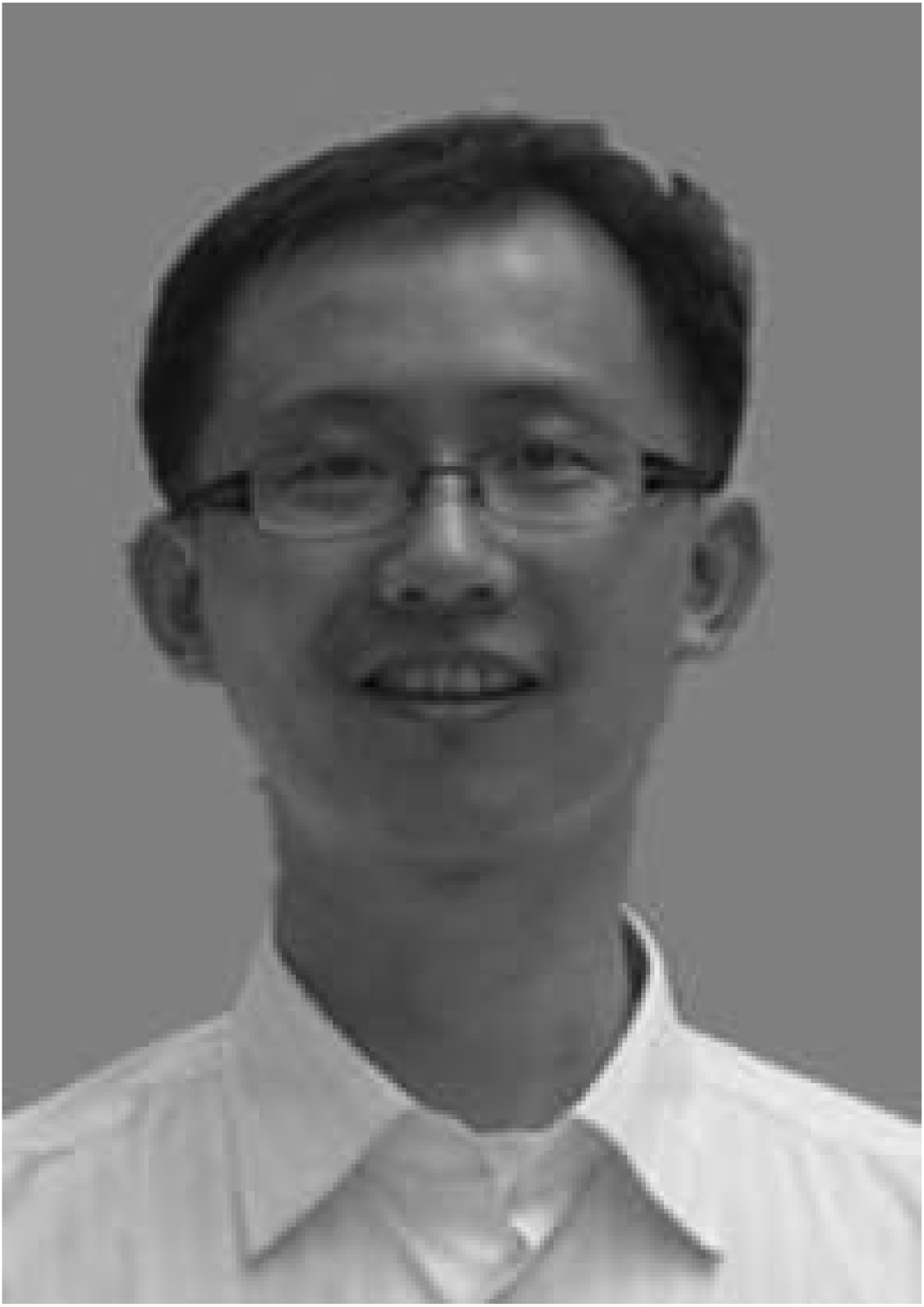}}]{Xiang Lian}
	received the BS degree from the Department of Computer Science and Technology, Nanjing University. in 2003, and the PhD degree in
	computer science from the Hong Kong University of Science and Technology, Hong Kong. He is now an assistant professor in the Department of
	Computer Science, Kent University. His research	interests include probabilistic/uncertain data management, 
	probabilistic RDF graphs, inconsistent probabilistic databases, and streaming time series.
\end{IEEEbiography}



\end{document}